\documentclass[12pt]{article}

\usepackage[utf8]{inputenc}
\usepackage{cancel}
\usepackage[normalem]{ulem}
\usepackage{amssymb}
\usepackage{soul}
\usepackage{tabularx}
\usepackage{xcolor}
\usepackage{booktabs}
\usepackage{subcaption} % Added by NN
\usepackage{colortbl}
\usepackage{authblk}
\usepackage[normalem]{ulem}
\usepackage{tablefootnote}
\DeclareUnicodeCharacter{00A0}{ }

\newcolumntype{C}{>{\centering\arraybackslash}X}
\usepackage[T1]{fontenc}
\usepackage{epsfig}
\usepackage{latexsym}
\usepackage{graphicx}
\usepackage{amsmath}
\usepackage{amsfonts}   % if you want the fonts
\usepackage{amssymb}    % if you want extra symbols
\usepackage{float}
\usepackage{bm}
\usepackage{url}
\usepackage{hyperref} 
\usepackage[nodisplayskipstretch]{setspace}
\def\lsim{\raise0.3ex\hbox{$\;<$\kern-0.75em\raise-1.1ex\hbox{$\sim\;$}}}
\def\gsim{\raise0.3ex\hbox{$\;>$\kern-0.75em\raise-1.1ex\hbox{$\sim\;$}}}
\newcommand{\captions}{\sf\caption}
\def    \beq            {\begin{equation}}
\def    \eeq            {\end{equation}}
\def    \bea           {\begin{eqnarray}}
\def    \eea           {\end{eqnarray}}

\def\g2{{\rm GeV}^2}
\def\sw2{sin^2 \theta_w}

\def\a^tau{\alpha_{\tau}}

\def\beq{\begin{equation}}
\def\eeq{\end{equation}}
\def\beqa{\begin{eqnarray}}
\def\eeqa{\end{eqnarray}}

\newcommand{\newc}{\newcommand}
\newc\BR{BR}
\newc{\akappa}{A_{\kappa} }
\newc\deltagmtwo{\delta (g-2)_{\mu}} 
\newc\deltaamu{\Delta a_{\mu}}

\def\anti{\overline}

\newc{\haa}{BR\(h_1\to a_1 a_1\)}
\newc{\abb}{BR\(a_1\to b\anti{b}\)}
\newc{\hbb}{BR\(h_1\to b\anti{b}\)}
\newc{\abund}{\Omega h^2}
\newc\bsgamma{b\rightarrow s \gamma }
\newc\bxsgamma{\overline{B}\rightarrow X_{s}\gamma}
\newc\brbsgamma{\BR(\overline{B}\rightarrow X_s\gamma)}

\newc{\Fermi}{\textit{Fermi}-}
\allowdisplaybreaks

\usepackage{array, makecell}
\usepackage{boldline}
\usepackage{cite}
\title{\bf{The new $(g-2)_\mu$ and Right-Handed Sneutrino Dark Matter
}}
\author[a]{Jong~Soo~Kim\thanks{jongsoo.kim@tu-dortmund.de}}
\author[b,c]{Daniel~E.~López-Fogliani\thanks{daniel.lopez@df.uba.ar}}
\author[d]{Andres~D.~Perez\thanks{andres.perez@iflp.unlp.edu.ar}}
\author[e]{Roberto~Ruiz~de~Austri\thanks{rruiz@ific.uv.es}}
\affil[a]{School of Physics, University of the Witwatersrand, Johannesburg, South Africa}
\affil[b]{Instituto de Física de Buenos Aires UBA \& CONICET, Departamento de Física,
Facultad de Ciencia Exactas y Naturales, Universidad de Buenos Aires, 
1428 Buenos Aires, Argentina}
\affil[c]{
{Pontificia Universidad Católica Argentina, 
Av. Alicia Moreau de Justo 1500, 
1107~Buenos~Aires, Argentina}}
\affil[d]{IFLP, CONICET - Dpto. de Física, Universidad Nacional de La Plata,\protect\\
C.C. 67, 1900 La Plata, Argentina}
\affil[e]{Instituto de Física Corpuscular CSIC-UV, c/Catedrático José Beltrán 2, 46980 Paterna (Valencia), Spain}

\renewcommand\Affilfont{\itshape\small}%Sets italic and size for affiliation
\date{}
\begin{document}
\maketitle
%%%%%%%%%%%%%%%%%%%%%%%%%%
%%%     Abstract    %%%%%%
%%%%%%%%%%%%%%%%%%%%%%%%%%
\begin{abstract}

In this paper we investigate the $(g-2)_\mu$ discrepancy in the context of the R-parity conserving next-to-minimal supersymmetric Standard Model plus right-handed neutrinos superfields. The model has the ability to reproduce neutrino physics data and includes the interesting possibility to have the right-handed sneutrino as the lightest supersymmetric particle and a viable dark matter candidate. 
%We find that our scenario satisfies all up to date constraints including \R{neutrino physics and} the latest results on $(g-2)_{\mu}$.
Since right-handed sneutrinos are singlets, no new contributions for $\delta a_{\mu}$ with respect to the MSSM and NMSSM are present.
%\sout{\R{at the 1-loop order}}. 
However, the possibility to have the right-handed sneutrino as the lightest supersymmetric particle %\bl{in the NMSSM+RHN} 
opens new ways to escape Large Hadron Collider and direct detection constraints. In particular, we find that dark matter masses within $10 \lesssim m_{\tilde{\nu}_{R}} \lesssim 600$~GeV are fully compatible with current experimental constraints. Remarkably, not only spectra with light sleptons are needed, but we obtain solutions with $m_{\tilde{\mu}} \gtrsim 600$~GeV in the entire dark matter mass range that could be probed by new $(g-2)_\mu$ data in the near future. In addition, %future 
dark matter direct detection experiments will be able to explore a sizable portion of the allowed parameter space with $m_{\tilde{\nu}_{R}} \lesssim 300$ GeV, while indirect detection experiments will be able to probe a much smaller fraction within $200 \lesssim m_{\tilde{\nu}_{R}} \lesssim 350$ GeV.

\end{abstract}
%%%%%%%%%%%%%%%%%%%%%%%%%%

%%%%%%%%%%%%%%%%%%%%%%%%%%
%%%      Keywords    %%%%%
%%%%%%%%%%%%%%%%%%%%%%%%%%


\newpage

%%%%%%%%%%%%%%%%%%%%%%%%%%%%%%%%%%%%%%%%%%%%%%%%%%%%%%%%%%%%%%%%%%%%%%%%%%%%
%%%%%%%%%%%%%%%%%%   Table of Contents          %%%%%%%%%%%%%%%%%%%%%%%%%%%%
%%%%%%%%%%%%%%%%%%%%%%%%%%%%%%%%%%%%%%%%%%%%%%%%%%%%%%%%%%%%%%%%%%%%%%%%%%%%
\tableofcontents 
%\listoffigures
%\listoftables
%%%%%%%%%%%%%%%%%%%%%%%%%%%%%%%%%%%%%%%%%%%%%%%%%%%%%%%%%%%%%%%%%%%%%%%%%%%%

%%%%%%%%%%%%%%%%%%%%%%%%%%%%%%%%%%%%%%%%%%%%%%%%%%%%%%%%%%%%%%%%%%%%%%%%%%%%
%%%%%%%%%%%%%%%%%%   Sections                   %%%%%%%%%%%%%%%%%%%%%%%%%%%%
%%%%%%%%%%%%%%%%%%%%%%%%%%%%%%%%%%%%%%%%%%%%%%%%%%%%%%%%%%%%%%%%%%%%%%%%%%%%

%%%%%%%%%%%%%%%%%%%%%%%%%%%%%%%%%%%%%%%%
\section{Introduction}
\label{sec:intro}
%%%%%%%%%%%%%%%%%%%%%%%%%%%%%%%%%%%%%%%%

There is an ongoing effort to search for physics beyond the Standard Model (BSM) of Particle Physics. Experiments looking for Dark Matter (DM) both directly and indirectly as well as searches for BSM at the Large Hadron Collider (LHC) have not found evidences for new physics so far. Nevertheless there are some experiments pointing to the existence of new physics. One of those is related with the anomalous magnetic moment of the muon $(g-2)_\mu$ whose measurements have a long standing discrepancy with the Standard Model (SM) predictions. 

The combination of the Fermilab-based Muon $(g-2)_\mu$ experiment measurement and the Brookhaven National Lab. (BNL) gives \cite{Abi:2021gix,Albahri:2021ixb}
\begin{equation}
a_{\mu}^{\text{exp}}=116 592 061(41) \times 10^{-11} \, ,
\end{equation}
where $a_\mu^{\text{exp}} \equiv \frac{(g-2)_\mu}{2}$.

On the other hand the SM prediction has been improved along the years, specially the hadronic vacuum polarization (HVP) from dispersive methods. The Muon (g-2) Theory Initiative recommended value is \cite{Aoyama:2020ynm},
\begin{equation}
a_{\mu}^{\text{SM}} = 116 591 810(43) \times 10^{-11} \, .
\end{equation}
It leads to 
\begin{equation}
\delta a_{\mu} \equiv a_{\mu}^{\text{exp}} - a_{\mu}^{\text{SM}} = (25.1 \pm 5.9) \times 10^{-10} \, , 
\end{equation}
corresponding to a 4.2 $\sigma$ deviation with the SM prediction. Indeed lattice QCD estimations were not included because, in general, they have a much larger uncertainties than the dispersive approaches which are data-driven \footnote{The exception is the results reported by the Budapest-Marseille-Wuppertal (BMW) collaboration on the calculation of the leading-order HVP which is compatible with the experimental result \cite{Borsanyi:2020mff}. However it has been shown that the BWM collaboration result is in tension with electroweak observables \cite{Crivellin:2020zul, Keshavarzi:2020bfy, deRafael:2020uif}.}.

Among all the proposed BSM theories, perhaps the most appealing one is supersymmetry (SUSY). In particular the minimal supersymmetric extension of the Standard Model (MSSM). One of the most interesting features of the model is that the lightest neutralino, when it is 
the lightest supersymmetric particle (LSP), can explain the DM abundance in the Universe that we have measured with experiments like Planck~\cite{Planck:2018vyg}. In addition, it has been shown to be able to explain the most recent estimation of the $(g-2)_\mu$ anomaly \cite{Endo:2021zal,  VanBeekveld:2021tgn,  Ahmed:2021htr, Athron:2021iuf, Chakraborti:2021bmv, Baer:2021aax,  Altmannshofer:2021hfu}. %\cite{Chakraborti:2021bmv, VanBeekveld:2021tgn}. 
This is shared with the minimal extension of the MSSM, called the next-to-minimal supersymmetric standard model (NMSSM)~\footnote{For reviews of this model see Ref.~\cite{Ellwanger:2010,Maniatis:2009re}.} which includes in its formulation a singlet superfield to solve dynamically the $\mu$-problem of the MSSM. The new $(g-2)_\mu$ result has also been discussed in the literature in the context of the NMSSM \cite{Abdughani:2021pdc, Cao:2021tuh}.
However, neither of these two models is able to reproduce neutrino observables~\cite{Pontecorvo:1969,FukudaSK:1998,Ahmed:2004,Araki:2005,Aliu:2005}.

To address this issue, the addition of right-handed (RH) neutrino superfields to the MSSM (MSSM+RHN)~\cite{Hall:1998,Arkani:2001,Faber:2019mti}, or the NMSSM (NMSSM+RHN)~\cite{Kitano:2000} allows to reproduce neutrino observables. Unlike in the MSSM extension, the latter framework allows extra terms in the superpotential between the singlet and the RH neutrinos superfields, $\lambda_N S N N$ and its corresponding soft-breaking term. This results in Majorana masses that do not have to be introduced by hand, but are dynamically generated at the electroweak scale after the singlet acquires VEV. Then, the mass of the active neutrinos is generated by a seesaw mechanism, with the neutrino Yukawa couplings of the same order as the electron Yukawa couplings.

Furthermore the SUSY partners of the RH neutrinos, the RH sneutrinos, can be an alternative candidate to the neutralinos for DM \cite{Cerdeno:2009,Deppisch:2008bp,Cerdeno2:2009,Demir:2010,Cerdeno:2011,Cerdeno2:2014,Cerdeno:2015,Cerdeno:2016,Chatterjee:2014,Cerdeno:2017sks,Ghosh:2019jzu,Cao:2019,Cao:2019ofo,Cao:2019qng}. In particular, in the NMSSM+RHN the aforementioned extra terms also play a crucial role in the RH sneutrino dark matter sector. The model presents a direct coupling between RH sneutrinos and Higgs bosons (through singlet mixing), and between RH sneutrinos and neutralinos (involving the terms $\lambda_N N N S$ and $\lambda S H H$). Hence, new decay and annihilation channels appear with respect to the MSSM+RHN, modifying the phenomenology significantly. There are new contributions to the thermal production of sneutrino relic
density allowing compatible DM solutions within a wide range of masses, specially for $10 \lesssim m_{\tilde{\nu}_{R}} \lesssim 100$~GeV. Additionally, these channels also open new avenues to explore the parameter space in DM experiments. Last but not least, the collider phenomenology of the NMSSM+RHN is similar to the MSSM because the lightest neutralino always decays into a RH sneutrino LSP and a neutrino and thus the resulting missing-$E_T$ (MET) is identical. However, the allowed mass hierarchies are different since the relative mass splitting between the lightest neutralino and next-to-lightest-neutralino is not fixed by DM constraints as in the MSSM.

Therefore, in this work we investigate the compatibility of the RH sneutrino DM scenario in the NMSSM+RHN with the new $(g-2)_\mu$ discrepancy together with other relevant constraints from LHC searches of new physics, Higgs and DM constrains as the relic abundance as well as from direct and indirect experiments.

Regarding $(g-2)_\mu$ in SUSY models its main contributions come from chargino-sneutrino and neutralino-smuon loops. One- and two-loop contributions in the MSSM are well-known and can be found in Ref.~\cite{Moroi:1995yh,Martin:2001st,Cerdeno:2001aj,Giudice:2012pf} and \cite{Degrassi:1998es,Heinemeyer:2004yq,Feng:2008cn,Fargnoli:2013zda,Fargnoli:2013zia,Athron:2015rva}, respectively. In some singlet extensions, like in the NMSSM, these contributions have the same expressions  provided that the mixing matrices are appropriately taken into account. 
%since the singlet and singlino mix with the Higgs and neutralinos and hence can be incorporated into the corresponding sectors.
Despite this, 
%redefinition,
different numerical values from the ones in the MSSM can be obtained~\cite{Gunion:2005rw,Domingo:2008bb}. On the other hand, in addition to the previously mentioned contributions, in the NMSSM+RHN new diagrams appear beyond the one-loop level involving RH sneutrinos and RH neutrinos since these mass states do not mix with the relevant MSSM sectors.
However these contributions are suppressed by small neutrino Yukawa coupling  $O(10^{-6})$, or
%However, they depend on the neutrino Yukawa values,
to be significant they would require the presence of an extremely light CP-odd scalar with large $\tan\beta$ and large mixing between the singlet and the $SU(2)$ components, or the need of and extremely light RH sneutrinos and large singlino-Higgsino mixing.
%The first type of contributions are suppressed by the smallness of the Yukawa couplings, $O(10^{-6})$ to reproduce the neutrino physics, the second is disfavored by Higgs measurements, and the third by the small singlino-Higgsino mixing found in our solutions with light DM. 
The second condition is disfavored by Higgs measurements, and the third by the small singlino-Higgsino mixing found in our solutions with light DM. 
Although a deeper study of $(g-2)_\mu$ two-loop corrections would be interesting, it is out of the scope of this work. We consider 
in this work
the leading one-loop contributions, denoted $\delta a_{\mu}$,
%throughout the rest of this work,
and focus on its impact on the distinctive features of the NMSSM+RHN highlighting the important differences with the MSSM and NMSSM phenomenology.

The paper is organised as follows. In Sec.~\ref{sec:model} we review the NMSSM+RHN model outlining the relevant sneutrino sector. In Sec.~\ref{sec:methodology} we describe the parameter scanning procedure and in Sec.~\ref{sec:results} we discuss our results finalizing in Sec.~\ref{sec:conclusions} with the conclusions.

%%%%%%%%%%%%%%%%%%%%%%%%%%%%%%%%%%%%%%%%
\section{Brief Description of the Model}
\label{sec:model}
%%%%%%%%%%%%%%%%%%%%%%%%%%%%%%%%%%%%%%%

In this work we follow the analysis done in \cite{Lopez-Fogliani:2021qpq}. The two candidates for DM in this context are the neutralino and the RH sneutrino. The relevant annihilation channels involved to achieve the correct amount of RH sneutrino relic density and the spin-independent sneutrino-nucleon scattering cross section  were discussed in \cite{Lopez-Fogliani:2021qpq} and references therein.

We summarize below the most relevant definitions and magnitudes in order to discuss our results in section \ref{sec:results}.

The superpotential and usual soft breaking terms are given by,  
\bea
W &=& \epsilon_{\alpha \beta} \left( Y_e^{ij} \, \hat{H}_d^{\alpha} \, \hat{L}_i^{\beta} \, \hat{e}_j \, + \, Y_d^{ij} \, \hat{H}_d^{\alpha} \, \hat{Q}_i^{\beta} \, \hat{d}_j \, + \, Y_u^{ij} \, \hat{Q}_i^{\alpha} \, \hat{H}_u^{\beta} \, \hat{u}_j \, + \, Y_N^{ij} \, \hat{L}_i^{\alpha} \, \hat{H}_u^{\beta} \, \hat{N}_j \, + \, \lambda \, \hat{S} \, \hat{H}_u^{\alpha} \, \hat{H}_d^{\beta} \right) \nonumber \\ & + & \lambda_N^{ij} \, \hat{N}_i \, \hat{N}_j \, \hat{S} \, + \, \frac{\kappa}{3} \, \hat{S}^3,
\label{WNMSSM}
\eea

\bea
V_{soft}&=& \biggl[ \epsilon_{\alpha\beta}  \, \biggl( A_e^{ij} \, Y_e^{ij} \, H_d^{\alpha} \, \tilde{L}_i^{\beta} \, \tilde{e}_j \, + \, A_d^{ij} \, Y_d^{ij} \, H_d^{\alpha} \, \tilde{Q}_i^{\beta} \, \tilde{d}_j \, + \, A_u^{ij} \, Y_u^{ij} \, \tilde{Q}_i^{\alpha} \, H_u^{\beta} \, \tilde{u}_j \, + \, A_N^{ij} \, Y_N^{ij} \, \tilde{L}_i^{\alpha} \, H_u^{\beta} \, \tilde{N}_j \nonumber \\ & + & A_{\lambda} \, \lambda \, S \, H_u^{\alpha} \, H_d^{\beta} \biggr) \, + \, A_{\lambda_N}^{ij} \, \lambda_N^{ij} \, \tilde{N}_i \, \tilde{N}_j \, S \, + \, \frac{A_{\kappa} \, \kappa}{3} \, S^3 \biggr] \, + h.c. \nonumber \\ & + & m^2_{\phi_{ij}} \, \phi_i^{\dagger} \, \phi_j \, + \, m^2_{\theta_{ij}} \, \theta_i \, \theta_j^* \, + \, m^2_{H_d} \, H_d^{\dagger} \, H_d \, + \, m^2_{H_u} \, H_u^{\dagger} \, H_u \, + \, m^2_S \, S \, S^* \nonumber \\ & + & \frac{1}{2} \, M_1 \, \tilde{B} \, \tilde{B} \, + \, \frac{1}{2} \, M_2 \, \tilde{W}^a \, \tilde{W}^a\, + \, \frac{1}{2} \, M_3 \, \tilde{g}^b \, \tilde{g}^b,
\label{VNMSSM}
\eea
where $\hat{S}$ is a singlet superfield, $\hat{N}$ the neutrino superfield, $\phi ={\tilde{L},\tilde{Q}}$; $\theta ={\tilde{e},\tilde{N},\tilde{u},\tilde{d}}$ are the scalar components of the corresponding superfields, and the gauginos $\tilde{B}, \tilde{W}, \tilde{g}$, are the fermionic superpartners of the $B$, $W$ bosons, and gluons. $i, j\in\{1,2,3\}$ are generation indices. $\alpha, \beta\in\{1,2\}$ are indices of the SU(2) fundamental representation while the $a\in\{1,2,3\}$ and $b\in\{1,\dots,8\}$ indices correspond to the adjoint representation of the SU(2) and SU(3), respectively.

As was done in \cite{Lopez-Fogliani:2021qpq} in this work we take all sfermion soft masses diagonal, $m^2_{ij}=m^2_{ii}=m^2_i$ and vanishing otherwise, were summation of repeated index convention was not used. Regarding the Yukawa and trilinear couplings,  only the third generation of sfermions are non-zero, $T^{ij}=A^{ij}Y^{ij}$, without the summation convention, except in the neutrino case where $Y_{N}^{ij}$ and $A_N^{ij}$ are taken diagonal. Furthermore, we also consider diagonal the parameter $\lambda_N^{ij}=\lambda_N^{ii}=\lambda_N^{i}$, and its corresponding trilinear coupling, $A_{\lambda_N}^{ij} \, \lambda_N^{ij}=A_{\lambda_N}^{i} \, \lambda_N^{i} = T_{\lambda_N}^{i}$.

After electroweak symmetry breaking (EWSB) induced by the soft SUSY-breaking terms of $O(\text{TeV})$, and with the choice of CP conservation, the neutral Higgses ($H_{u,d}$) and the scalar singlet develop the following vacuum expectation values (VEVs)
\begin{equation}
\langle H_d \rangle = \frac{v_d}{\sqrt{2}}, \hspace{1cm} \langle H_u \rangle = \frac{v_u}{\sqrt{2}}, \hspace{1cm} \langle s \rangle = \frac{v_s}{\sqrt{2}},
\label{vevs}
\end{equation}
where $v^2=v_d^2+v_u^2= 4m_Z^2 /(g^2 + g'^2) \simeq (246 \text{ GeV})^2$, with $m_Z$ the $Z$ boson mass, and $g$ and $g'$ the $U(1)_Y$ and $SU(2)_L$ couplings, correspondingly. 

In appendix \ref{appendix} we make a summary of used notation and the mass matrices for CP-even and CP-odd Higgses, as well as for charged Higgses, charginos and neutralinos. In the next subsection we briefly discuss the sneutrino sector crucial for our discussion.

%%%%%%%%%%%%%%%%%%%%%%%%%%%%%%%%%%%%%%%%
\subsection{Sneutrino sector}
\label{sec:sneutrinosector}
%%%%%%%%%%%%%%%%%%%%%%%%%%%%%%%%%%%%%%%%

Since we consider the RH sneutrino as the DM particle in this model, in order to continue our discussion we  summarize in this subsection the $12\times 12$ mass matrix for the sneutrino sector divided into $6 \times 6$ submatrices,
\begin{equation}
M_{\tilde{\nu}}^2 = 
\left(\begin{array}{cc}
 m_{\mathbb{R} \mathbb{R}}^2 & 0_{6\times 6}\\
  0_{6\times 6} & m_{\mathbb{I} \mathbb{I}}^2
  \end{array}\right),
\end{equation}
where the subscripts $\mathbb{R}$ and $\mathbb{I}$ denote CP-even and CP-odd states, respectively. The off-diagonal submatrices are zero due to our choice of CP conservation. The submatrices are
\begin{equation}
m_{\mathbb{R} \mathbb{R}}^2 = 
\left(\begin{array}{cc}
 m_{L_{i}}^2 & A_{i}^{+}\\
  (A_{i}^{+})^T & m_{R_{i}}^2 + B_i
  \end{array}\right),  \hspace{1cm}  m_{\mathbb{I} \mathbb{I}}^2 = 
\left(\begin{array}{cc}
 m_{L_{i}}^2 & A_{i}^{-}\\
  (A_{i}^{-})^T & m_{R_{i}}^2 - B_i
  \end{array}\right),
\end{equation}
with
\bea
A_i^{+} &=& Y_N^{i} \, \left( \, A_N^i \, v_u \, + \, 2 \, \lambda_N^{i} \, v_u \, v_s \, - \, \lambda \, v_d \, v_s \, \right),\label{aplus}\\
A_i^{-} &=& Y_N^{i} \, \left( \, A_N^i \, v_u \, - \, 2 \, \lambda_N^{i} \, v_u \, v_s \, - \, \lambda \, v_d \, v_s \, \right),\label{aminus}\\
B_i &=& 2 \, \lambda_N^{i} \, \left( \, A_{\lambda_N}^{i} \, v_s \, + \, \kappa \, v_s^{2} \, - \, \lambda  \, v_u \, v_d \, \right),\label{bsneutrino}\\
m_{L_i}^2 &=& m_{\tilde{L}_i}^2 \, + \, (Y_N^{i})^2 \, v_u^2 \, + \, \frac{1}{2} \, m_Z^2 \, \cos 2 \beta ,\\
m_{R_i}^2 &=& m_{\tilde{N}_i}^2 \, + \, (Y_N^{i})^2 \, v_u^2 \, + \, 4 \, (\lambda_N^{i})^2 \, v_s^2,
\eea
here the index $i,j = 1,2,3$ are the family indices. The mixing between left-handed (LH) and RH sneutrinos is suppressed by the small neutrino Yukawa value in Eq.~(\ref{aplus}) and (\ref{aminus}).

As was discussed in \cite{Lopez-Fogliani:2021qpq}, considering only one family of neutrinos the mixing angle between LH and RH sneutrinos, $\theta_{\tilde{\nu}}$, for typical parameter values, $A_N \sim A_{\lambda_N} \sim O(\text{GeV})$, $\tan \beta \sim O(10)$, $\lambda \sim \kappa \sim \lambda_N$, can be approximated by,
\bea
\tan 2 \theta_{\tilde{\nu}} & \sim & \frac{ Y_N \, \lambda_N \, v_s \, v_u}{m_{\tilde{L}}^2 \, + \, \frac{1}{2} \, m_Z^2 \, \cos 2 \beta \, - \, m_{\tilde{N}}^2 \, - \, \lambda_N^2 \, v_s^2 } \sim 10^{-2} \times Y_N \sim O(10^{-8}), \label{sneutmix2}
\eea
where we have used that $m_{\tilde{L}} \sim O(10^3)$ GeV to evade the stringent collider constraints on SUSY particles, and that $\lambda_N \, v_s \sim O(\text{EW})$ with $Y_N\sim 10^{-6}$ to reproduce the neutrino masses (see appendix~\ref{appendixB}).

Due to the small mixing, the RH sneutrino masses can be taken as
\bea
m_{\tilde{\nu}_{R_i}}^2 \simeq m_{R_i}^2 \, \pm \, B_i,
\eea
here also the upper (lower) sign corresponds to the CP-even (CP-odd) state. We can see that the mass splitting is proportional to $\lambda_N^i$ in Eq.~(\ref{bsneutrino}). If $\lambda_N \rightarrow 0$ then $m_{\tilde{\nu}_{R_i}}^2 \simeq  m_{\tilde{N}_i}^2$. For typical parameter values in the NMSSM plus RH neutrinos
\bea
m_{\tilde{\nu}_{R_i}}^2 \approx m_{\tilde{N}_i}^2 + (2 \, \lambda_N^{i} \, v_s)^2 \pm \left( 2 \, T_{\lambda_N}^{i} \, v_s \, + \, 2 \, \lambda_N^{i} \, \kappa \, v_s^{2} \right).
\label{RHsneutrinomassAprox}
\eea
The new parameters with respect to the NMSSM $m_{\tilde{N}}^2$, $\lambda_N$ and $T_{\lambda_N}$ can be chosen to set the physical RH sneutrino mass, without affecting the rest of the mass spectrum. Moreover, the last two parameters also determine the RH sneutrino coupling to the singlet Higgs boson.

%%%%%%%%%%%%%%%%%%%%%%%%%%%
\section{Scan Methodology}
\label{sec:methodology}
%%%%%%%%%%%%%%%%%%%%%%%%%%%

\begin{table}
\begin{small}
\begin{center}
\begin{tabular}{|c|c|}
        \hline
        \textbf{Constraint} & \textbf{} \\
        \hline
        \hline
            $\delta a_{\mu}$ & $(25.1 \pm 5.9) \times 10^{-10}$~\cite{Abi:2021gix} \\
            DM relic density ($\Omega_{cdm}^{\text{Planck}}h^2$) & Planck~\cite{Planck:2018vyg} \tablefootnote{In the following sections we highlight the solutions that satisfy Planck constraint on DM relic density ($0.1198 \pm 0.0012$) within 3$\sigma$. However, we also keep solutions with lower relic density allowing the possibility of multiple DM candidates.} \\
            SI cross section ($\sigma^{\text{SD}}_{DM-p}$) & XENON1T~\cite{xenon1tcite1,xenon1tcite2} \\
            Annihilation cross section ($\langle \sigma_{DM} v\rangle$) & \Fermi-LAT~\cite{Ackermann:2015dwarfs} and H.E.S.S.~\cite{HESS:2016} \\
            Higgs constraints & LEP, Tevatron, and LHC (\texttt{HiggsBounds}~\cite{HBcite1,HBcite2,HBcite3}) \\
            Higgs signal & LHC and Tevatron (\texttt{HiggsSignals}~\cite{HScite1,HScite2,HScite3}) \\
            $BR(b \rightarrow s \gamma)$ & $(3.27 \pm 0.14) \times 10^{-4}$~\cite{Misiak:2017} \\
            $BR(B_s \rightarrow \mu^+ \mu^-)$ & $(2.8^{+0.8}_{-0.7}) \times 10^{-9}$~\cite{ATLASBflavor:2019} \\
            $BR(\mu \rightarrow e \gamma)$ & $< 4.2 \times 10^{-13}$~\cite{MEGmuegamma:2016} \\
            $BR(\mu \rightarrow eee)$ & $< 1.0 \times 10^{-12}$~\cite{SINDRUMmueee:1988} \\
            R-hadrons (long-lived colored particles) & LHC~\cite{ATLASllcharged:2019} \\
            Chargino masses & $m_{\tilde{\chi}^{\pm}_1} > 103.5$ GeV~\cite{Zyla:2020zbs}\\
            Other collider constraints & LHC (\texttt{SModelS}~\cite{Kraml:2013mwa})\\
        \hline
\end{tabular}
\end{center}
\end{small}
  \caption{Constraints that have been applied to our model set (see text for details).}
  \label{TableConstraints}
\end{table}

As aforementioned, in this work we focus on finding solutions in which the RH sneutrino is the LSP and a good DM candidate in addition to fulfill the phenomenological constraints summarized in Table \ref{TableConstraints}. Notice that we consider Planck constraint on DM relic density as an upper limit for our DM candidate, allowing the possibility of multiple DM candidates in extensions of the model, for instance including axions.
%, e.g. axions.}

In order to compare the model predictions with experimental constraints we have generated a version of \texttt{SPheno}~\cite{sphenocite1,sphenocite2} of the model using \texttt{SARAH} ~\cite{sarahcite1,sarahcite2,sarahcite3}. With \texttt{SPheno} we compute the spectrum as well as flavour observables and $\delta a_{\mu}$. \texttt{MicrOmegas}~\cite{micromegascite1,micromegascite2, micromegascite3} is used to compute the DM relic density, the present annihilation cross section ($\langle \sigma_{DM} v\rangle$), and the spin-independent WIMP-nucleon scattering cross section\footnote{The sneutrino is a scalar field, i.e. no axial-vector coupling in the effective Lagrangian, then it has vanishing spin-dependent cross section.} ($\sigma^{\text{SI}}_{DM-p}$). Besides,  \texttt{HiggsBounds}~\cite{HBcite1,HBcite2,HBcite3} and \texttt{HiggsSignals}~\cite{HScite1,HScite2,HScite3} are used to determine whether the SUSY models satisfy LEP, Tevatron and LHC Higgs constraints and measurements. The impact of DM direct detection experiments is computed with \texttt{DDCalc}~\cite{ddcalccite} during the scan.

With respect to LHC constraints we use \texttt{SModelS} \cite{Kraml:2013mwa} to impose 8 TeV and 13 TeV new physics searches performed by the ATLAS and CMS experiments. \texttt{SModelS} maps theoretical models onto simplified models LHC signatures and therefore can be directly compared with the existing LHC bounds if there is a match in the respective topologies.

The starting point to find viable solutions is Ref.~\cite{Lopez-Fogliani:2021qpq}, where we performed a scan using \texttt{MultiNest} to sample the parameter space of the model without including $\delta a_\mu$ in the combined likelihood. We denote this dataset as \emph{Scan 1}.

The ranges of the parameters used are shown in the second column labeled as \emph{Scan 1} in Table \ref{scanparameters} and for the rest we assumed universality for soft-sfermion masses $m^2_{\tilde{e}_i}=m^2_{\tilde{L}_i}=m^2_{\tilde{d}_i}=2.25\times 10^{6}$ GeV$^2$ with $i=1,2,3$, and $m^2_{\tilde{N}_i}=m^2_{\tilde{u}_i}=m^2_{\tilde{Q}_i}=2.25\times 10^{6}$ GeV$^2$ with $i=1,2$. In addition we set $T_d^{33}=T_{d_3}=256$ GeV, and $T_e^{33}=T_{e_3}=-98$ GeV while $Y_N^{i}=10^{-6}$. Besides we considered vanishing $T_{N}^i=A_N^i \, Y_N^i$ as an approximation due to the small neutrino Yukawa couplings and thus we took $A_{_N}^i\sim O(\text{GeV})$. To simplify the analysis, we set $\lambda_N^i=-0.5$, and $T_{\lambda_N}^{i} = 0$ for $i=1,2$, to obtain two families of heavy sneutrinos.
Finally we fixed $M_3=3$ TeV.

\begin{table}[t]
      \centering
        \hspace{0.9cm}\begin{tabular}{|c|c|c|}
        \hline
        \textbf{Parameter} & \textbf{Scan 1} & \textbf{Scan 2} \\
        \hline
        \hline
         & \\[-2ex]
            $M_1$ & (20, 3000) GeV & (40, 1500) GeV \\[1ex]
            $M_2$ & (20, 3000) GeV & (100, 1500) GeV \\[1ex]
            $\mu_{eff}$ & (100, 5000) GeV & $\mu_{eff}^0\pm$10\% \\[1ex]
            $\tan \beta$ & (2, 50) & ($\tan \beta^0-10\%$, 50) \\[1ex]
            $\lambda$ & (0.001, 0.8) & $\lambda^0\pm$10\% \\[1ex]
            $\kappa$ & (0.001, 0.8) & $\kappa^0\pm$10\% \\[1ex]
            $\lambda_N^3$ & (-0.4, -0.001) & $\lambda_N^{3^0}\pm$10\% \\[1ex]
            $T_{\lambda}$ & (0.001, 600) GeV & $T_{\lambda}^0$ \\[1ex]
            $T_{\kappa}$ & (-30, -0.001) GeV & $T_{\kappa}^0$ \\[1ex]
            $T_{\lambda_N}^3$ & (-1100, -0.001) & $T_{\lambda_N}^{3^0}$ \\[1ex]
            $m^2_{\tilde{N_3}}$ & ($10$, $2.5\times10^{6}$) GeV$^2$ & $m^2_{\tilde{N_3}^0}\pm$10\% \\[1ex]
            $m^2_{\tilde{u}_3}$ & ($2.5\times10^{5}$, $4\times10^{6}$) GeV$^2$ & $m^2_{{{\tilde{u}_3}}^0}$ \\[1ex]
            $m^2_{\tilde{Q}_3}$ & ($2.5\times10^{5}$, $4\times10^{6}$) GeV$^2$ & $m^2_{{\tilde{Q}_3}^0}$ \\[1ex]
            $m^2_{\tilde{L}_1}$ = $m^2_{\tilde{L}_2}$ & $2.25\times10^{6}$ GeV$^2$ & (($\mu_{eff}^0-50 \text{GeV}$)$^2$, $1\times10^{6}$) GeV$^2$ \\[1ex]
            $T_{u_3}$ & (700, 10000) GeV & $T_{u_3}^0$ \\[1ex]
        \hline
        \end{tabular}
   \caption{Sampling ranges and fixed parameters used in our scan. The superscript `0' refers to the seed value of the corresponding parameter found in \emph{Scan 1}.}
  \label{scanparameters}
\end{table}

The main contributions to the muon anomalous magnetic moment involve LH smuons and neutralinos. We would like to highlight that due to our previous choice of $m^2_{\tilde{L}_2}=2.25\times 10^{6}$ GeV$^2$ most solutions found in the \emph{Scan 1} dataset present low values of $\delta a_{\mu}$. In this work we use those solutions with RH sneutrino DM as seeds, and perform several scans varying nine parameters in subranges: 
\begin{equation}
    M_1, \: M_2, \: \lambda_N, \: m^2_{\tilde{N}_{3}}, \: \kappa, \: \mu_{eff}, \: \lambda, \: \tan \beta,  \: m_{L_2},
\end{equation}
whose values can be seen on the third column of Table~\ref{scanparameters}, labeled as \emph{Scan 2}.

To fulfil $\delta a_{\mu}$ constraints, we take $m_{L_1} = m_{L_2}$ within $[\mu^{0}_{eff} - 50 \: \text{GeV}, 1000 \: \text{GeV}]$, where superscript `0' refers to the seed value of $\mu_{eff}$.
We consider $M_1 = [40, 1500]$ GeV, and $M_2 = [100, 1500]$ GeV, to get lighter neutralinos with different compositions.
To obtain different DM masses (see
Eq.~(\ref{RHsneutrinomassAprox})), we allow variations up to $\pm$10\% with respect to the seed values of $\lambda_N$, $m^2_{\tilde{N}_{3}}$, and $\kappa$. As we have pointed out in Sec.~\ref{sec:intro}, % shown in Ref.~\cite{Lopez-Fogliani:2021qpq},
the existence of direct couplings of RH sneutrino to Higgs bosons and neutralinos through terms involving $\lambda_N$ is a crucial feature of this model. This makes RH sneutrino DM viable, allowing annihilation mechanisms that involve resonances to CP-even Higgs, direct annihilations to pseudo-scalar particles, and coannihilations with neutralinos. Therefore, to satisfy the Higgs sector constraints for solutions with different RH sneutrino masses we also scan the parameters $\mu_{eff}$, $\lambda$, and $\tan \beta$ within a $\pm$10\% range from their previously obtained values. In the case of $\tan \beta$ we set its upper bound to 50, in order to allow solutions with higher $\delta a_{\mu}$. 

In appendix~\ref{appendixB} we show the values of $\mu_{eff}$, $v_s$, and $Y_N$ for the solutions found with the \emph{Scan 2} dataset. For the Yukawa couplings that satisfy the neutrino physics we get $m_{\nu}\sim O(0.1 \text{ eV})$.

%%%%%%%%%%%%%%%%%%%
\section{Results}
\label{sec:results}
%%%%%%%%%%%%%%%%%%

\begin{figure}[t!]
\begin{center}
 \begin{tabular}{c}
 \hspace*{-10mm}
 \epsfig{file=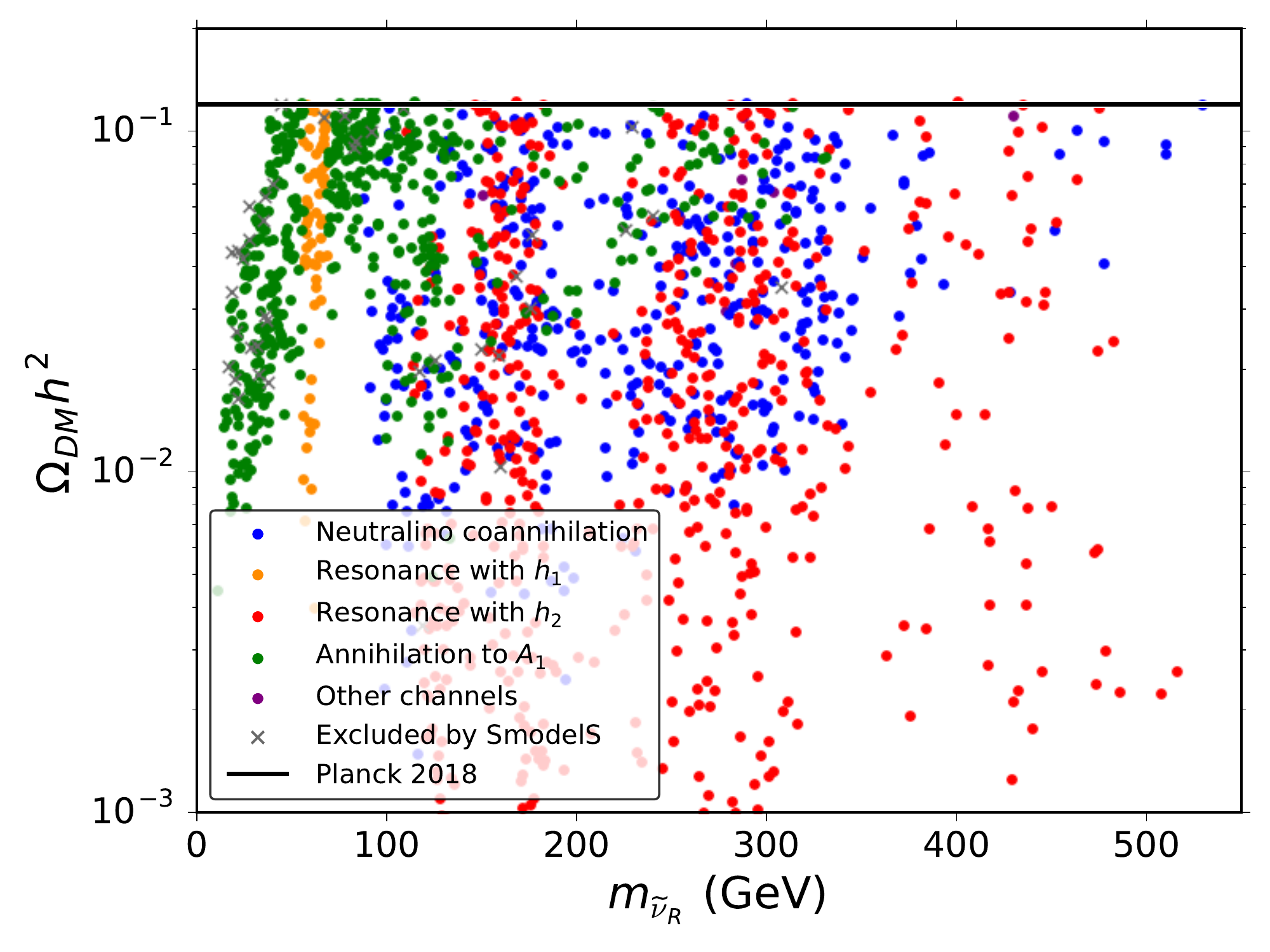,height=9cm} 
 
    \end{tabular}
    \captions{Relic density for the parameter points with a RH sneutrino as DM candidate that fulfill all the constraints considered in this work. The color coding represents the main channels used by RH sneutrinos to obtain an allowed relic density. The black solid line corresponds to the amount of DM measured by the Planck Collaboration~\cite{Planck:2018vyg}.
}
    \label{fig_relic}
\end{center}
\end{figure}

%In this section we discuss the results starting for
We start the discussion of our results by showing in Fig.~\ref{fig_relic} the relic density as a function of the RH sneutrino DM mass for the solutions that fulfill the constraints considered in this work. To be conservative, we allow points with $\delta a^\mu$ compatible within $3 \sigma$, since one has to account for the theoretical uncertainty in the evaluation of $\delta a^\mu$ (i.e. two-loop corrections are not included in our computation.). However, we would like to highlight that points with values of $\delta a^\mu$ within 1 or 2$\sigma$ can be found in the entire mass range, and can also explain the total amount of DM relic abundance measured by the Planck Collaboration~\cite{Planck:2018vyg}, shown as a black solid line.

The color coding represents the main annihilation mechanisms used by RH sneutrinos to obtain an allowed relic density. Namely:
\begin{itemize}
    \item Funnel with the lightest (orange) 
    and second lightest (red) CP-even scalar,
    \item RH sneutrinos self-annihilating with a four-vertex to a lightest pseudo-scalar pair~(green),
    \item Coannihilation with lightest neutralinos (blue).
\end{itemize}

For $m_{\tilde{\nu}_R} \lesssim 100$ GeV, the dominant processes are resonances with the lightest CP-even scalar and annihilations via direct quartic couplings to the lightest pseudo-scalar. In this mass region almost no coannihilations can be found due to the lower bound imposed by LEP-II on the chargino mass ($m_{\tilde{\chi}^{\pm}_1} \gtrsim 103.5$ GeV).

On the other hand, for $m_{\tilde{\nu}_R} \gtrsim 100$ GeV, the dominant annihilation mechanisms are coannihilations with neutralinos, mostly through the Higgsino component as RH sneutrino couplings with neutralinos involve the terms $\lambda_N N N S$ and $\lambda S H H$ (see Eq.~(\ref{WNMSSM})). Since usually the lightest CP-even scalar is a SM-like Higgs boson, a small singlet component for the SM-like Higgs is required. At the same time, very important contributions come from resonances with the second lightest CP-even scalar (usually singlet dominated), and annihilations via direct quartic couplings. For the former mechanism, solutions with very low relic density values can be found. 

For $m_{\tilde{\nu}_R}\gtrsim 500$ GeV, finding solutions becomes an increasingly challenging task. As expected, $\delta a^\mu$ tends to decrease for heavy SUSY particles.

%Notice that a crucial difference of this model with respect to the MSSM or NMSSM is that RH sneutrino as DM is not involved in the main contribution channels to $\delta a^\mu$. In that sense, the DM mass range is not constrained directly by this parameter. \bl{In the MSSM and NMSSM the DM candidate is usually a neutralino whose mass window is pushed from below by collider searches and from above by $\delta a^\mu$ measurements. In the NMSSM+RHN the DM mass range can be wider since scenarios with very light RH sneutrino LSP, $m_{\tilde{\nu}_R} \lesssim m_Z / 2$, can escape LHC searches as long as the mass spectrum contains a heavier neutralino-chargino sector to provide significant contributions to $\delta a^\mu$ while satisfying collider bounds.} Nonetheless, notice that if the RH sneutrino employs coannihilations mechanism with neutralinos to achieve an allowed relic density, the mass splitting should be small $\lesssim 10 \%$. In this case, the allowed sneutrino mass range will inherit the constraints on the neutralino (and chargino) sector regarding its masses.

%\R{
Notice that a key characteristic of the model is that the DM candidate considered, the RH sneutrino, is not involved in the main contribution channels to $\delta a^\mu$. In that sense, the DM mass in the NMSSM+RHN is not constrained directly by the $(g-2)_\mu$ measurements. Moreover, we can have scenarios with a very light DM candidate, i.e. $m_{\tilde{\nu}_R} \lesssim m_Z / 2$. This is not the case for the MSSM with a neutralino LSP. Although its contribution to $\delta a^\mu$ can be small depending on the region of parameter space (see for example Ref.~\cite{Endo:2021zal}), the neutralino sector is affected directly by both DM and $\delta a^\mu$ measurements.
Nonetheless, notice that if the RH sneutrino employs a coannihilation mechanism with neutralinos to achieve an allowed relic density, the mass splitting should be small $\lesssim 10 \%$. In this case, the allowed sneutrino mass range will inherit the constraints on the neutralino (and chargino) sector regarding its masses. 
%}

\begin{figure}[t!]
\begin{center}
 \begin{tabular}{c}
 \hspace*{-10mm}
 \epsfig{file=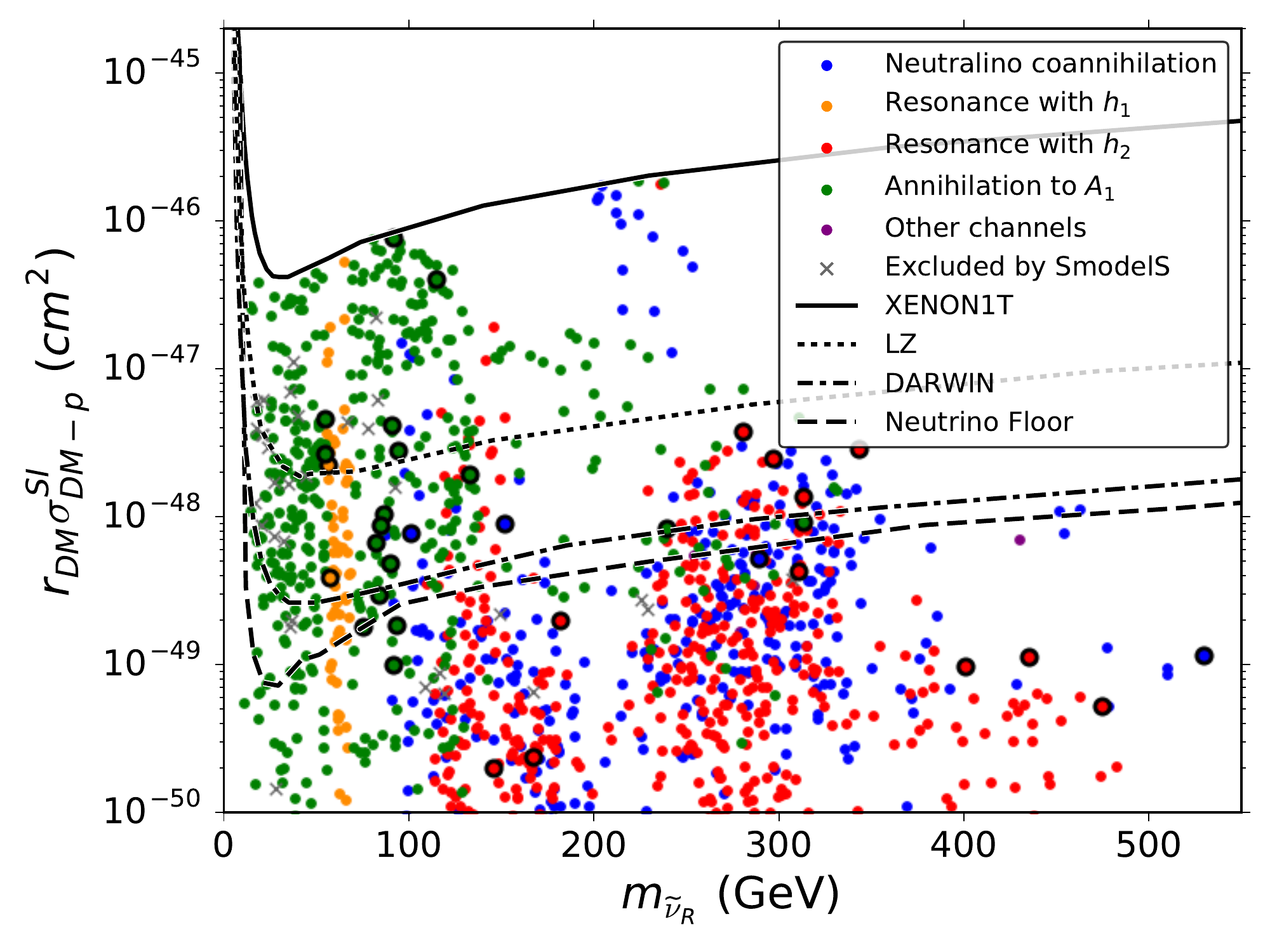,height=9cm} 

    \end{tabular}
    \captions{Scaled spin-independent direct detection cross section vs RH sneutrino mass. The color coding represents the main channels used by RH sneutrinos to obtain an allowed relic density, points with $\Omega_{DM}h^2$ satisfying Planck constrain within 3$\sigma$ are marked with a black edge. The solid black curve shows the current experimental sensitivities from XENON1T~\cite{xenon1tcite1,xenon1tcite2}. Projected sensitivities for LZ~\cite{LZ:2015kxe} and DARWIN~\cite{DARWIN:2016} experiments are shown as black dotted and dot-dashed curves, respectively. The black dashed curve shows the neutrino floor taken from Ref.~\cite{Billard:2014,Ruppin:2014}. The scaling factor $r_{DM}$ accounts for the possibility that the calculated thermal relic density lies below the Planck measurement.
}
    \label{fig_cross_section}
\end{center}
\end{figure}

In Fig.~\ref{fig_cross_section} we show the predicted scaled spin-independent (SI) scattering cross sections of the RH sneutrino with a proton for the allowed points of our scan. The same color coding as in Fig.~\ref{fig_relic} is used. The scaling factor $r_{DM}=\frac{\Omega_{DM}h^2}{\Omega_{cdm}^{\text{Planck}}h^2}$ allows to compare solutions with relic density below the Planck measurement against published constraint. We also present the current experimental constraints from XENON1T~\cite{xenon1tcite1,xenon1tcite2}, the projected sensitivities from LZ~\cite{LZ:2015kxe} and DARWIN~\cite{DARWIN:2016}, and the neutrino background floor~\cite{Billard:2014,Ruppin:2014}.

If the only relevant scattering proceeds by exchanging the lightest Higgs boson $h_1$ that is SM-like, then we can estimate~\cite{Lopez-Fogliani:2021qpq}
\begin{equation}
    \sigma^{SI}_{\widetilde{\nu}_R-p}\; \approx \; \frac{(F_u^{(p)})^2}{\pi} \frac{m_p^4}{m_{\widetilde{\nu}_R}^2m_{h_1}^4} \left( \frac{\lambda \; \lambda_N}{\tan \beta} \right)^2.
    \label{sigmaSIapprox1}
\end{equation}
Solutions with RH sneutrino DM using coannihilations with neutralinos prefer low values of $|\lambda_N^3|$ and $\lambda$, therefore tend to produce very small SI cross section values.

As can be seen from the figure, the points that would be explored by LZ and DARWIN have low DM masses, mostly $m_{\tilde{\nu}_R} \lesssim 150$
GeV, and therefore, they do not coannihilate with neutralinos. The dominant mechanism in this region involves annihilations to a pair of pseudo-scalars through direct quartic coupling, a channel not present in the MSSM+RHN. Resonances with a CP-even Higgs scalar are also relevant. 
As expected, for increasing DM masses it is more difficult to obtain solutions above the neutrino floor, specially for $m_{\tilde{\nu}_R} \gtrsim 350$ GeV.

\begin{figure}[t!]
\begin{center}
 \begin{tabular}{c}
 \hspace*{-10mm}
 \epsfig{file=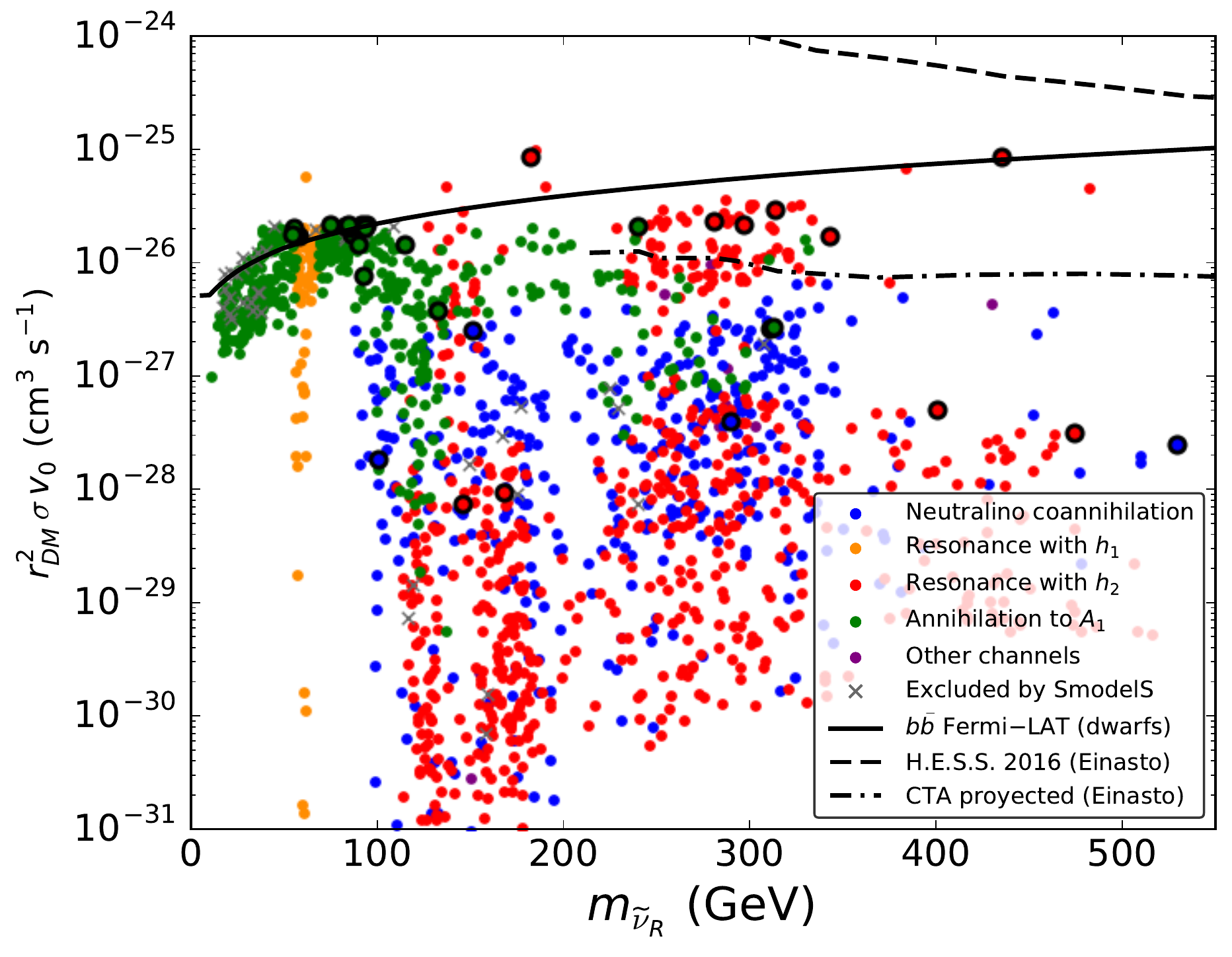,height=9cm} 

    \end{tabular}
    \captions{Solutions in the ($m_{DM}$, $r_{DM}^2 \, \langle \sigma v_0\rangle$) space, where $r_{DM}$ is the DM relic density fraction. The color coding represents the main channels used by RH sneutrinos to obtain an allowed relic density, points with $\Omega_{DM}h^2$ satisfying Planck constrain within 3$\sigma$ are marked with a black edge. The current upper 95\% C.L. limits from \Fermi-LAT to $b \bar{b}$ from the observation of dwarfs~\cite{Ackermann:2015dwarfs}, and the limits of H.E.S.S. from observations of the Galactic center using Einasto profile~\cite{Hryczuk:2019} are indicated as solid and dashed black curves, respectively. The projected CTA sensitivity~\cite{Hryczuk:2019} is shown as a dot-dashed black curve.}
    \label{fig_indirect_detection}
\end{center}
\end{figure}

Regarding indirect detection experiments, in Fig.~\ref{fig_indirect_detection}, we show the normalized thermally averaged annihilation cross section, $r_{DM}^2 \times \langle \sigma v_0\rangle$, as a function of the DM mass, where we have used the scaling factor $r_{DM}$ as in the direct detection case. Upper bounds on annihilation cross sections are derived for pure channels and applied at face value. %because direct detection limits are usually more restrictive.
Since we obtain a mixture of annihilation channels, the data sets were implemented on each of the reported channels. If an annihilation final state has not been reported, we employ the most relevant existing bounds. 

To guide the eye the solid black curve shows the limit of $b \bar{b}$ final state set by \Fermi-LAT from the observation of dwarfs galaxies~\cite{Ackermann:2015dwarfs}. The dashed black curve represents the annihilation lower limits from observations of the Galactic center by H.E.S.S. collaboration~\cite{HESS:2016} using Einasto profile calculated in Ref.~\cite{Hryczuk:2019}. The dot-dashed black curve corresponds to the projected CTA~\cite{CTAConsortium:2017dvg} 95\% C.L. sensitivity to DM annihilation derived from observations of the Galactic center assuming 500 hour homogeneous exposure and Einasto profile~\cite{Hryczuk:2019}. Unlike direct detection prospects of detection, the small fraction of points that will be explored by CTA mostly present funnels with the second lightest CP-even scalar as the dominant annihilation channel, and masses up to $\sim 350$ GeV.

On the other hand, to explore RH sneutrino DM that coannihilates with neutralinos we have to rely on collider studies, as neither direct nor indirect detection experiments will be able to probe them in the near future.

\begin{figure}[t!]
\begin{center}
 \begin{tabular}{cc}
 \hspace*{-14mm}
 \epsfig{file=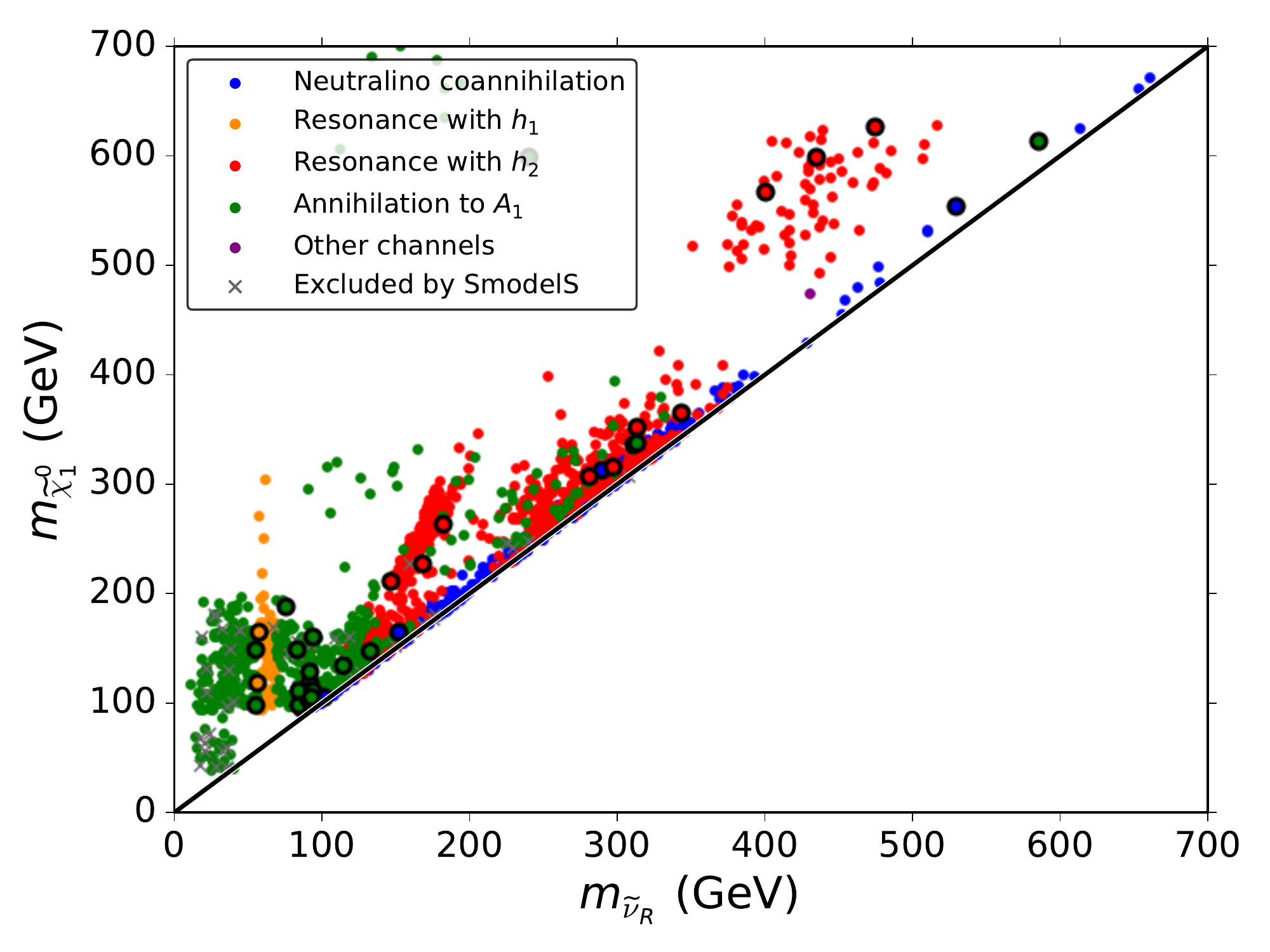,height=6.9cm} 
        \hspace*{-3mm}\epsfig{file=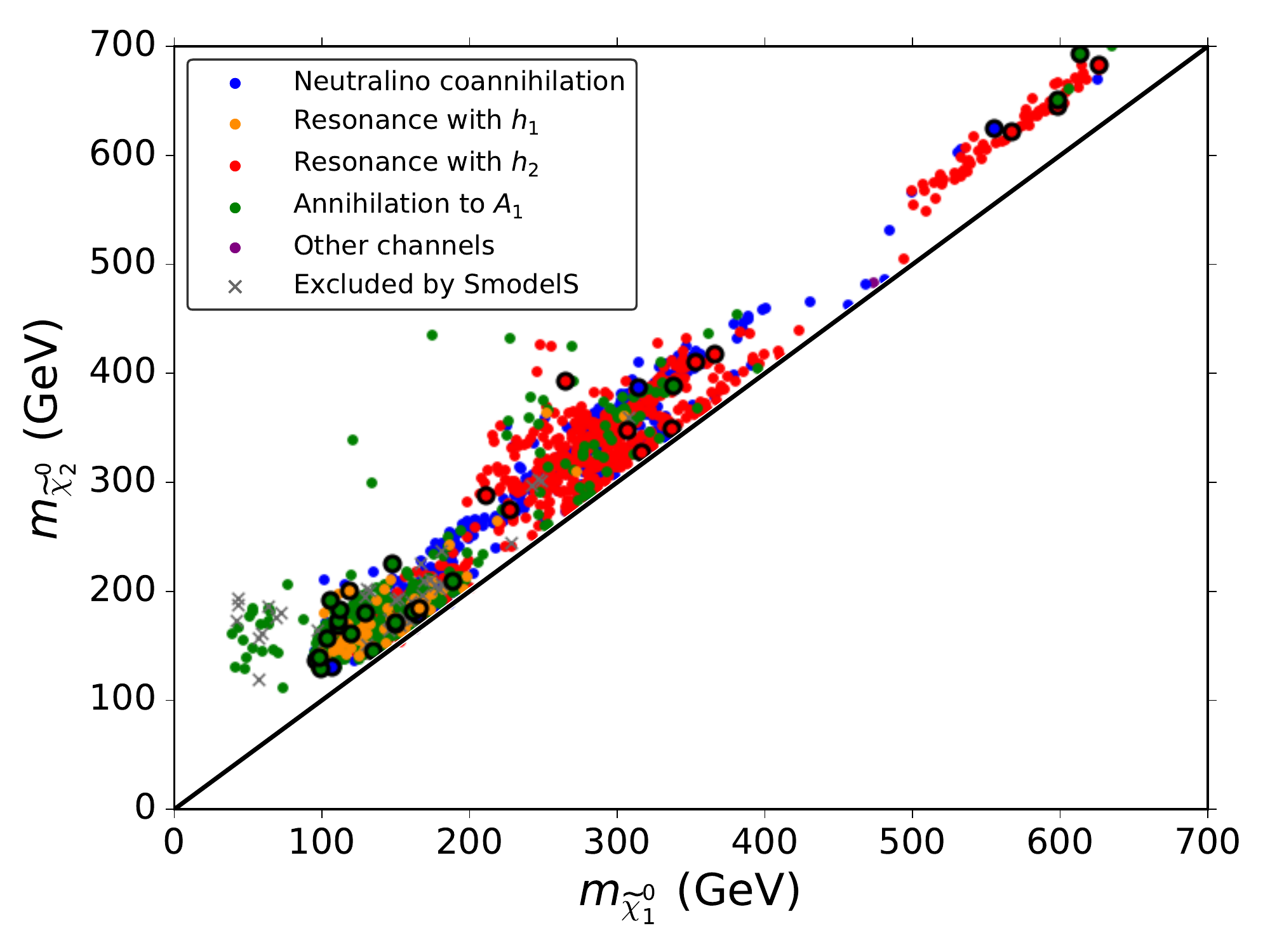,height=6.9cm} \\
 \hspace*{-14mm} \epsfig{file=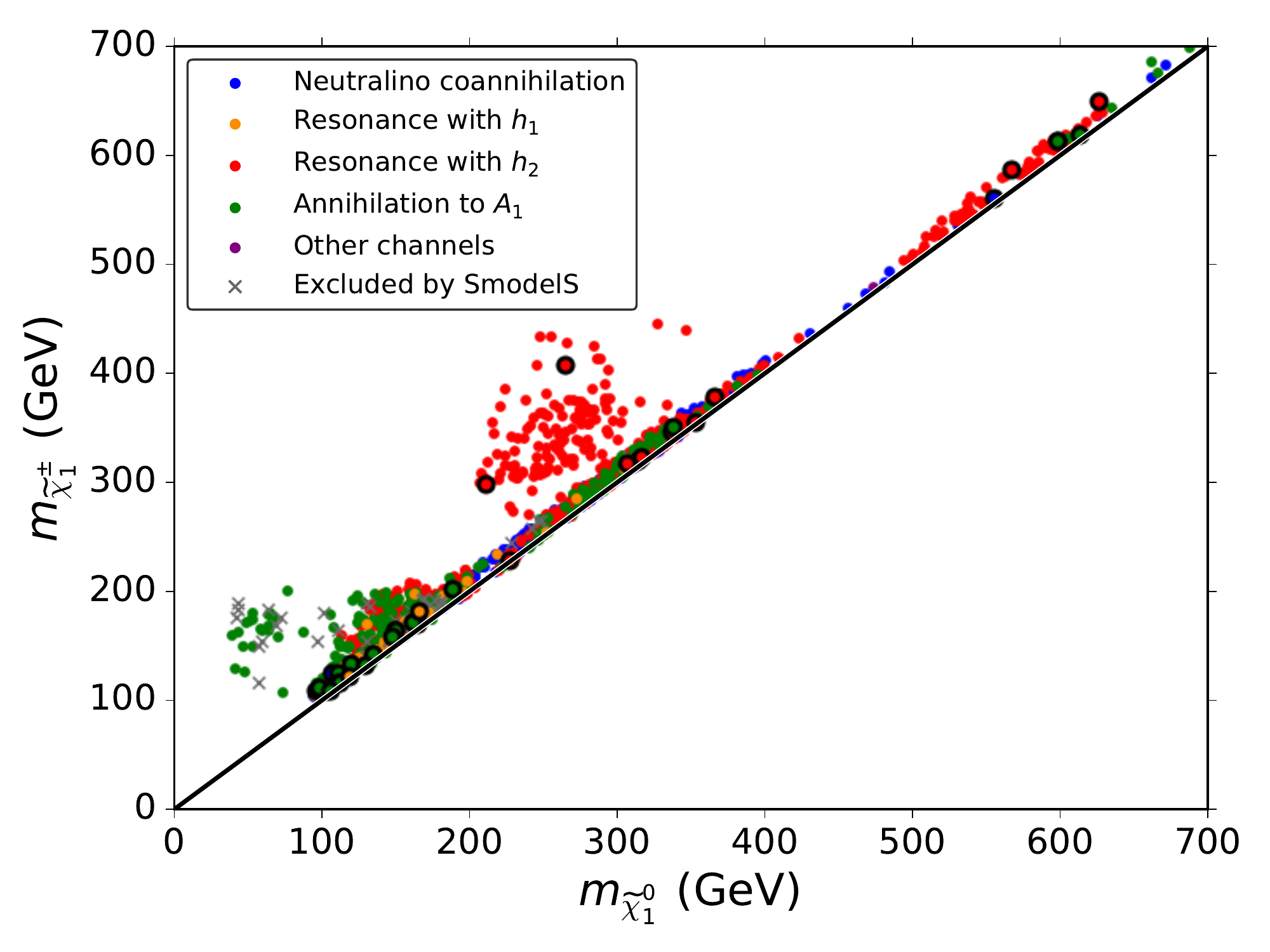,height=6.9cm} 
        \hspace*{-3mm}\epsfig{file=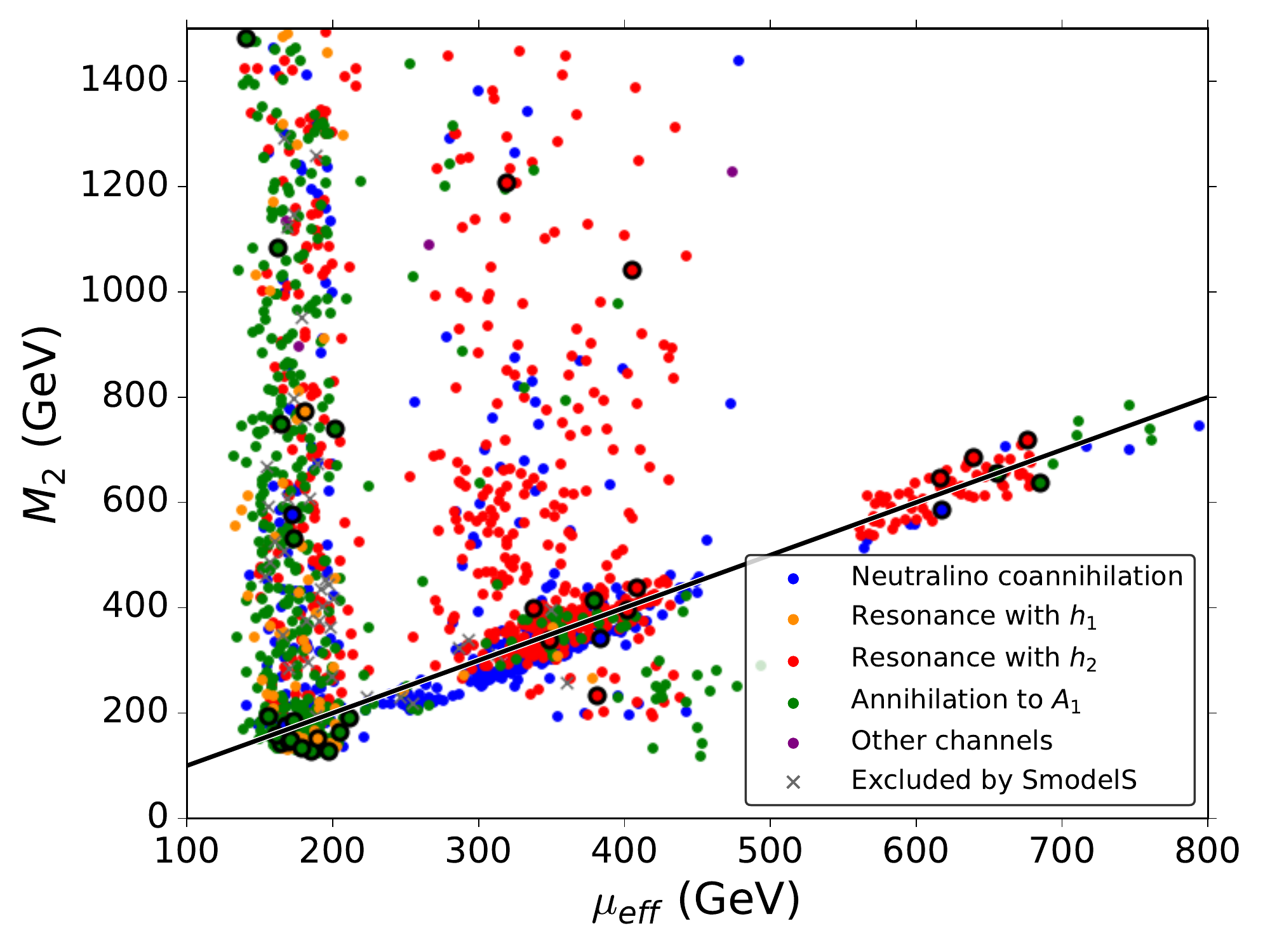,height=6.9cm}  \\
    \end{tabular}
    \captions{Relevant SUSY spectrum. The points color coding is the same as in Fig.~\ref{fig_relic}, and points with $\Omega_{DM}h^2$ satisfying Planck constrain within 3$\sigma$ are marked with a black edge.
}
    \label{spectrum1}
\end{center}
\end{figure}

\begin{figure}[t!]
\begin{center}
 \begin{tabular}{cc}
 \hspace*{-14mm} \epsfig{file=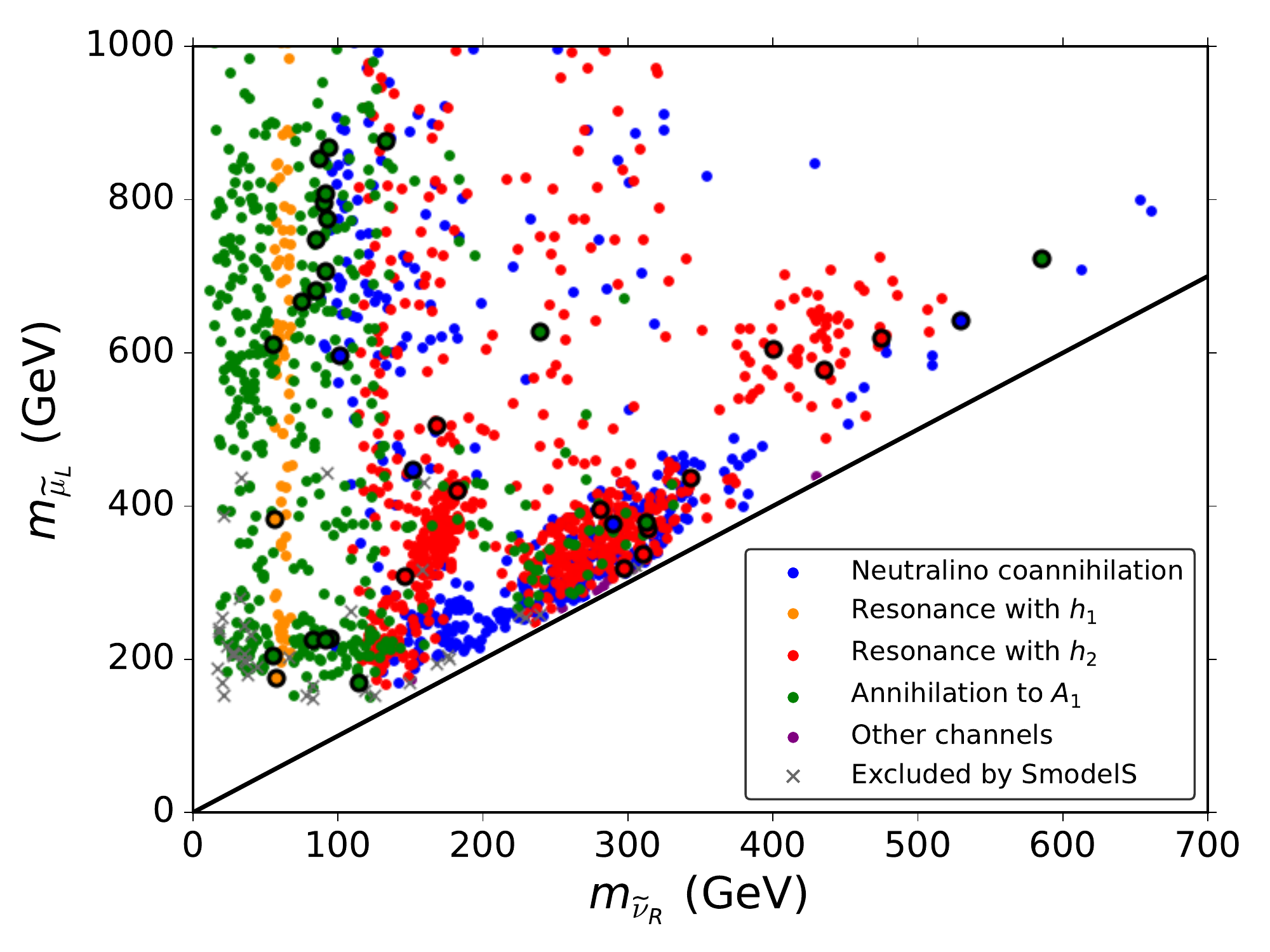,height=6.9cm} 
        \hspace*{-3mm}\epsfig{file=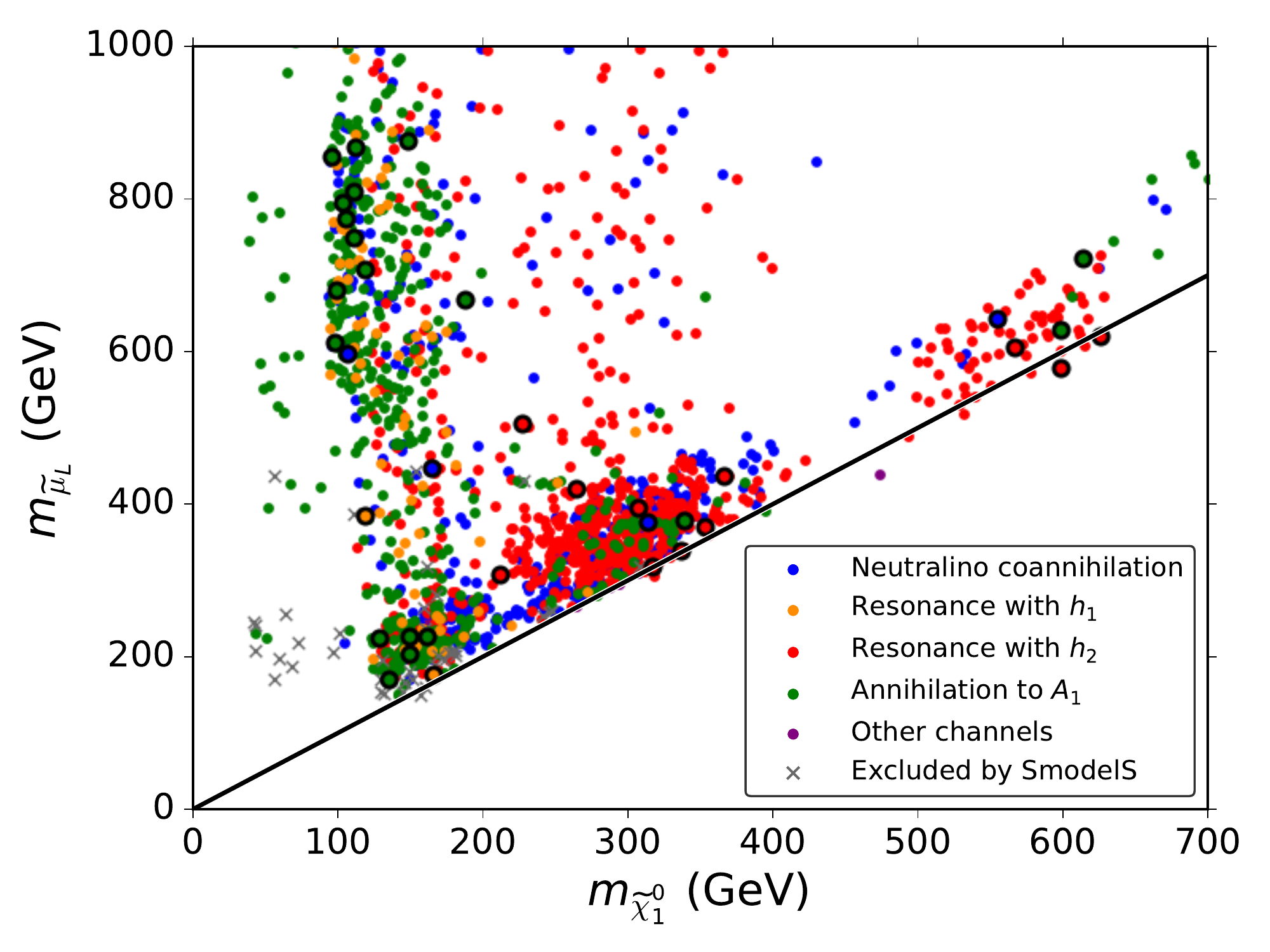,height=6.9cm}   
    \end{tabular}
    \captions{Same as in Fig.~\ref{spectrum1}. Notice that $m_{\tilde{\mu}_L} \simeq m_{\tilde{\nu}_{\mu L}}$ as they belong to the same $SU(2)$ doublet.
}
    \label{spectrum2}
\end{center}
\end{figure}

%%%%%%%%%%%%%%%%%%%%%%%%%%%%%%%%%%%%%%%%%%%%%%
\subsection{Interplay $g-2$ with LHC signatures}
%%%%%%%%%%%%%%%%%%%%%%%%%%%%%%%%%%%%%%%%%%%%%%
In this section we briefly discuss the LHC constraints. Since we have decoupled the strongly interacting supersymmetric sector we expect that only electroweakino and slepton searches are relevant for our study.\footnote{In principle, long lived particle searches might be relevant. However, we explicitely checked that only prompt searches are important.} ATLAS \cite{Aad:2021ajl,Aad:2019vnb, Aad:2019vvi, Aad:2019qnd} and CMS \cite{CMS:2021few, CMS:2021cox} have presented a large number of electroweakino and slepton searches which are interpreted in the simplified models framework, however, a priori we cannot directly exploit their published limits because they make very specific assumptions on the exact nature of the supersymmetric particles. In particular, in our solutions the LSP is not the lightest neutralino which is the assumption in most dedicated searches of both collaborations. In our scenario, the lightest supersymmetric particle is always the RH sneutrino.

On the other hand, the collider phenomenology is very similar to the MSSM since even though our lightest supersymmetric particle is not the neutralino the final missing transverse momentum of our model will be identical to the MSSM. The reason is that the NLSP is always the lightest neutralino decaying into a neutrino and sneutrino without exception in our benchmark points.\footnote{There could be a slight twist to the story. In our model a chiral singlet superfield with a singlino and a singlet scalar is present. If light enough, the heavier (higgsino-like) neutralino eigenstates could decay into them and thus possibly alter our collider signatures compared to the MSSM case \cite{Domingo:2018ykx}. However, we checked that our benchmark points are indeed MSSM like.} However, the allowed mass hierarchies, and hence the DM phenomenology, are different compared to the MSSM since the relative mass splitting between the neutralino NLSP and the next-to-NLSP is not fixed by DM constraints. The allowed parameter space in the MSSM requires the mass splitting between the neutralino LSP and the NLSP to be small enough so that coannihilation leads to the correct relic density. This is not true for our model. As a result, even taking DM constraints into account, non compressed collider signatures are found. In addition, in our model the MET distributions measured at the LHC cannot be directly used to deduce the mass of the DM candidate compared to the MSSM.

In this work, a detailed LHC study is beyond the scope of this work. Instead, we want to discuss the collider constraints with the recasting tool \texttt{SModelS 2.0.0}~\cite{Kraml:2013mwa}.

In Fig.~\ref{spectrum1} we present the collider constraints in various mass planes.
We keep the color labels of the previous figures but tag benchmark points with grey colored cross labels if excluded by \texttt{SModelS}.
The top-left panel displays the $\tilde{\chi}_1^0$ - $\tilde{\nu}_R$ mass plane. As argued before, the lightest neutralino is always the NLSP and it decays into the RH sneutrino and the corresponding neutrino and as result the final missing transverse momentum is fixed by the lightest neutralino mass and its relative mass difference to the heavier supersymmetric particles.

The top-right and bottom-left panels show the masses in the $\tilde\chi_2^0$ - $\tilde\chi_1^0$ and $\tilde\chi_1^\pm$ - $\tilde\chi_1^0$ planes. We see that the $\tilde\chi_2^0$ and $\tilde\chi_1^\pm$ masses are mostly \textit{aligned}. However, we also observe that in some regions the mass degeneracy between $\tilde\chi_2^0$ and $\tilde\chi_1^\pm$ is absent. In the bottom-right panel we show our benchmark points in the $M_2$ - $\mu$ plane. Here, we see that mostly light higgsino states and a few mixed wino-higgsino states are excluded. Our findings are roughly consistent with the results found in Ref.~\cite{Adhikary:2020ujn,Ambrogi:2017lov}. We also see that we do not have many wino dominated eigenstates. At closer inspection, it turns out that the few lighter wino dominated eigenstates are almost excluded by SModels.

In Fig.~\ref{spectrum2}, we display the $\tilde\mu_L$ - $\tilde\nu_R$ and $\tilde\mu_L$ - $\tilde\chi_1^0$ mass planes. There we can see that most of the excluded points have $m_{\tilde{\mu}_L} \lesssim 400$ GeV. Since the LH sneutrinos are in the same $SU(2)$ doublet as the charged LH sleptons, both are mass degenerate $m_{\tilde{\mu}_L} \simeq m_{\tilde{\nu}_{\mu L}}$. Pair produced LH sneutrinos can result in a dilepton and transverse missing energy final state that shares the same constraints as for charged sleptons.

The main conclusion drawn from this section is that a large fraction of benchmark points with light electroweakinos and/or sleptons are still allowed even after taking into account the constraints from the full Run 2 data. A few benchmark points with chargino/neutralino masses and slepton masses up to about 300 GeV and 400 GeV can be excluded, respectively. However, it should be stressed that the derived collider limits are very conservative and more accurate constraints can only be obtained by using dedicated recasting tools like CheckMATE \cite{Drees:2013wra,Dercks:2016npn,Desai:2021jsa,Kim:2015wza}, ColliderBit \cite{Balazs:2017moi} or MadAnalysis 5 \cite{Conte:2012fm}.

\begin{figure}[t!]
\begin{center}
 \begin{tabular}{cc}
 \hspace*{-14mm} \epsfig{file=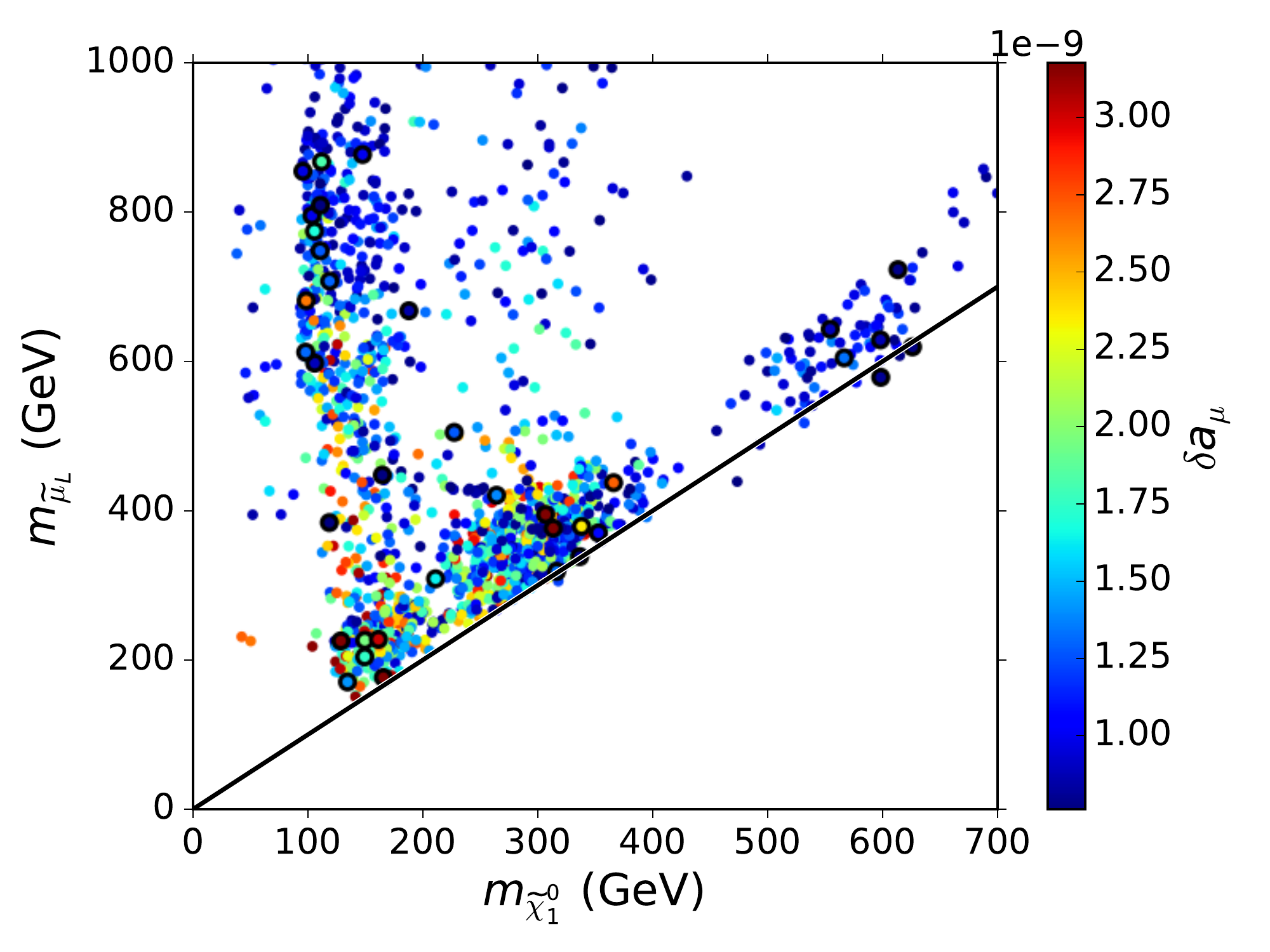,height=6.9cm} 
        \hspace*{-3mm}\epsfig{file=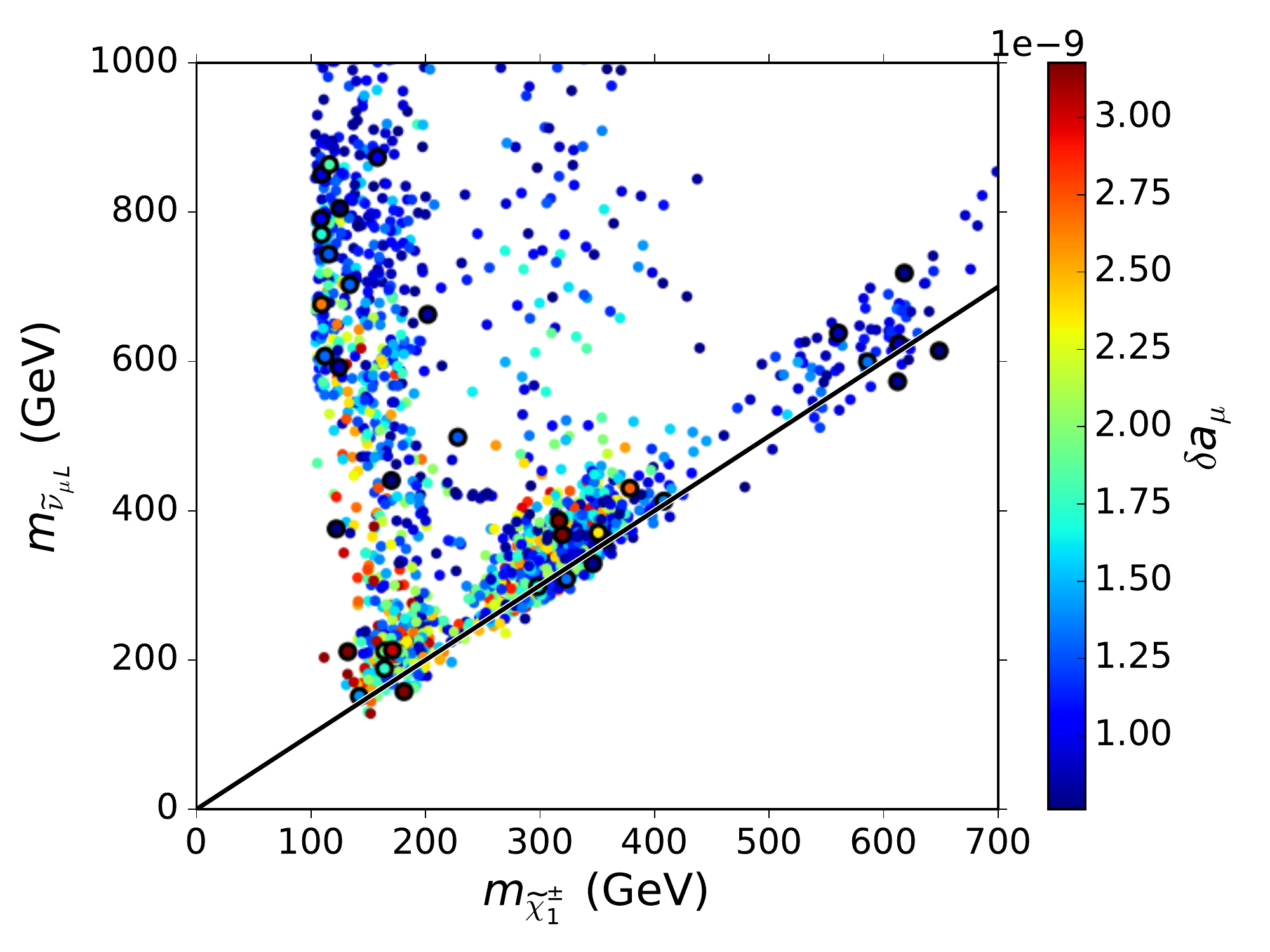,height=6.9cm}   
    \end{tabular}
    \captions{Masses of the particles involved in the main contribution channels to $(g-2)_{\mu}$. Points with $\Omega_{DM}h^2$ satisfying Planck constrain within 3$\sigma$ are marked with a black edge.
}
    \label{spectrum3g2}
\end{center}
\end{figure}

To summarize our results in Fig.~\ref{spectrum3g2} we show the correlation between $(g-2)_{\mu}$ and the masses of the particles that are involved in its dominant contributions. These points satisfy current DM and LHC constraints, and satisfy the recent anomalous magnetic moment of the muon measurement. While the former ones tend to disfavor low-mass sparticles, the latter pushes the mass limit from above. As expected, increasing the sparticle masses decreases $\delta a_{\mu}$, narrowing down the allowed mass range. This complementarity hints a promising path for upcoming experiments.

%%%%%%%%%%%%%%%%%%%
\section{Conclusions}
\label{sec:conclusions}
%%%%%%%%%%%%%%%%%%

In this work we have analyzed the impact of recent measurements of the muon anomalous magnetic moment in the context of the NMSSM including RH neutrino superfields requiring that the RH sneutrino is the LSP. Within this context, we have shown that the reported 4.2$\sigma$ deviation with respect to the SM prediction can be naturally explained.

To explore the model and to identify regions with viable RH sneutrino DM, we have scanned the parameter space employing \texttt{Multinest} as optimizer. We have imposed several constraints, including the measured amount of DM in the Universe, direct and indirect detection experiments of DM, neutrino physics, Higgs searches at colliders as well as other constraints on SUSY particles at the LHC. Model points consistent with experimental data with sleptons and electroweak gauginos with masses of order 100 GeV can be found, which is known to enhance the SUSY contribution to $(g-2)_\mu$. Interestingly a significant amount of points within constraints present heavy sleptons.

One key feature of the model is the existence of RH sneutrino direct couplings to Higgs bosons and to neutralinos, absent in both the NMSSM and the MSSM plus RH neutrinos. This extra terms are crucial to obtain the correct relic density, and to allow small but measurable DM-nucleon scattering and DM annihilation cross section in a wide mass range.
Distinctive annihilation channels are characteristic of RH sneutrino, including resonances, annihilations through scalar quartic couplings and coannihilations with all kind of neutralinos.
Funnel with the lightest CP-even Higgs and quartic coupling dominate the low mass spectrum, while funnels with the second lightest CP-even Higgs and coannihilations with mainly Higgsinos are dominant for heavy DM. 
We found that $10 \lesssim m_{\tilde{\nu}_{R}} \lesssim 600$~GeV, 
although for $m_{\tilde{\nu}_{R}} \gtrsim 350$~GeV 
it is increasingly harder to get solutions satisfying $(g-2)_{\mu}$.
Unlike in the neutralino DM case, where the LSP is directly involved in $(g-2)_\mu$ channels, the RH sneutrino is a singlet and therefore can cover the mentioned low mass region and escape some regions that would otherwise be disfavored, specially for $m_{\tilde{\nu}_{R}} \lesssim 100$~GeV.

We would also like to mention that the exploration of the parameter space relies on the complementarity of several searches. In that regard, if a signal is detected in both collider and dedicated DM experiments, different LSP masses could be derived, since a singlet RH sneutrino LSP would produce the same missing transverse momentum signature as the neutralino NLSP that decays to invisible. Direct detection DM experiments will be able to explore a sizable region of the parameter space with low RH sneutrino masses, $m_{\tilde{\nu}_{R}} \lesssim 300$ GeV, employing scalar quartic couplings with the lightest pseudo-scalar as the dominant annihilation mechanism. However indirect detection prospects are  weaker since they will be available to probe a much smaller portion of the parameter space. Namely DM masses within $200 \lesssim m_{\tilde{\nu}_{R}} \lesssim 350$ GeV corresponding to points using  funnels with the second lightest CP-even Higgs to annihilate. Even though, some RH sneutrino DM coannihilating with neutralinos could be explored with the mentioned DM experiments, collider constraints are important for this and the other annihilation mechanisms, specially for low mass SUSY particles. Also, if the $(g-2)_\mu$ deviation with respect to the SM increases, the mass range of RH sneutrino DM would narrow down, as heavy solutions are increasingly harder to obtain.

%\pagebreak

%%%%%%%%%%%%%%%%%%%%%%%%%%%%%%%%%%%%%%%%
%%%%%%%%%%%%%%%%%%%%%%%%%%%%%%%%%%%%%%%%

\section*{Acknowledgments}

The work of DL and AP was supported by the Argentinian CONICET, and also acknowledges the support through PIP11220170100154CO. They would like to thank the team supporting the Dirac High Performance Cluster at the Physics Department, FCEyN, UBA, for the computing time and their dedication in maintaining the cluster facilities. 
R. RdA acknowledges partial funding/support from the Elusives ITN (Marie Sklodowska-Curie grant agreement No 674896), the ``SOM Sabor y origen de la Materia" (FPA 2017-85985-P). JSK thanks Florian Domingo and Suchita Kulkarni for useful comments.

%%%%%%%%%%%%%%%%%%%%%%%%%%%%%%%%%%%%%%%%
%%%%%%%%%%%%%%%%%%%%%%%%%%%%%%%%%%%%%%%%

\pagebreak
\appendix
\section{Mass Matrices and useful formulas}
\label{appendix}

In this appendix we summary our notation and useful mass matrices formulas.\\

\noindent {\bf Scalars}

The scalar components of the superfields $\hat{H_u}$, $\hat{H_d}$, and $\hat{S}$ can be written as
\bea
H_u=\left(\begin{array}{c}
 H_u^+\\
  \frac{v_u}{\sqrt{2}} + \frac{H_u^{\mathbb{R}} \, + \, i \, H_u^{\mathbb{I}}}{\sqrt{2}} 
  \end{array}\right),  \hspace{1cm}      H_d=\left(\begin{array}{c}
  \frac{v_d}{\sqrt{2}} + \frac{H_d^{\mathbb{R}} \, + \, i \, H_d^{\mathbb{I}}}{\sqrt{2}} \\
  H_d^-
  \end{array}\right),  \hspace{1cm}     S=\frac{v_s}{\sqrt{2}} + \frac{S^{\mathbb{R}} \, + \, i \, S^{\mathbb{I}}}{\sqrt{2}},
\eea
where the superscripts $\mathbb{R}$ and $\mathbb{I}$ indicate CP-even and CP-odd component fields, respectively.

In the basis ($H_d^{\mathbb{R}}$, $H_u^{\mathbb{R}}$, $S^{\mathbb{R}}$), we denote the CP-even scalar mass matrix as $M_S^2$. Likewise, dropping off the Goldstone mode, in the basis ($A$, $S^{\mathbb{I}}$), with $A=H_u^{\mathbb{I}} \cos \beta + H_d^{\mathbb{I}} \sin \beta$ and $\tan \beta = \frac{v_u}{v_d}$, the CP-odd scalar mass matrix is denoted $M_P^2$. The mass eigenstates of the CP-even Higgs $h_i$ with $i=1,2,3$, and the CP-odd Higgs $A_i$ with $i=1,2$ can be obtained by
\bea
\left(\begin{array}{c}
 h_1\\
 h_2\\
 h_3  
  \end{array}\right) = S_{ij} \left(\begin{array}{c}
 H_d^{\mathbb{R}}\\
 H_u^{\mathbb{R}}\\
 S^{\mathbb{R}} 
  \end{array}\right),      \hspace{2cm}     \left(\begin{array}{c}
 A_1\\
 A_2  
  \end{array}\right) = P_{ij} \left(\begin{array}{c}
 A\\
 S^{\mathbb{I}}
  \end{array}\right),
  \label{cphiggsmatrices}
\eea
where the matrices $S_{ij}$ and $P_{ij}$ diagonalize the mass matrices $M_S^2$ and $M_P^2$, respectively. The states are labeled according to the mass hierarchy $m_{h_1} < m_{h_2} < m_{h_3}$, and $m_{A_1} < m_{A_2}$. We can estimate $m_{A_i} \propto A_{\kappa} \, \kappa \, v_s$
for the singlet dominated boson, which is important to allow RH sneutrino annihilations to light CP-odd scalars.\\

\noindent {\bf Neutrinos}

To generate an effective $\mu$-term, $\mu_{eff}=\frac{\lambda \, v_s}{\sqrt{2}}$, $v_s=O(\text{GeV-TeV})$ is needed. Then, the superpotential term $\lambda_N^{i} \, \hat{N}_i \, \hat{N}_i \, S$ generates dynamically a RH neutrino Majorana mass term $M_{N}^{i}=\frac{\lambda_N^{i} \, v_s}{\sqrt{2}} \sim O(\text{GeV-TeV})$, assuming that the parameters $\lambda$ and $\lambda_N^{i}$ are $O(0.1 - 1)$.

The LH neutrino masses, $m_{\nu_{L}}$, are generated by a seesaw mechanism. The general neutrino mass matrix is given by
\begin{equation}
M_{\nu} = 
\left(\begin{array}{cc}
 0 & m_D\\
  m_D^T & M_N
  \end{array}\right),
\end{equation}
therefore,
\bea
m_{\nu_{L}}\simeq - m_D \, M_N^{-1} \, m_D^T, 
\eea
with the dirac mass $m_D \simeq \frac{Y_N \, v_u}{\sqrt{2}}$. Thus $m_{\nu_{L}} \simeq \frac{Y_N^2 \, v_u^{2}}{\lambda_N \, v_s} \sim Y_N^2 \; \times$ EW scale, which implies $Y_N \sim 10^{-6}$ to get neutrino masses within the right order of magnitude. This estimation is enough for our discussion since the specific structure of the $Y_N$ does not affect our results.\\

\pagebreak
\noindent {\bf Neutralinos and charginos}

The neutralino and chargino sectors are the same as in the NMSSM. The neutral colorless gauginos mix with the neutral higgsinos-singlinos and generate a symmetric 5$\times$5 mass matrix $M_{\chi^0}$. In the basis ($-i\tilde{B}$, $-i\tilde{W}^3$, $\tilde{H}_d^0$, $\tilde{H}_u^0$, $S$) we get
\bea
M_{\chi^0} = 
\left(\begin{array}{ccccc}
 M_1 & 0 & -\frac{g \, v_d}{2} & \frac{g \, v_u}{2} & 0\\
  0 & M_2 & \frac{g' \, v_d}{2} & -\frac{g' \, v_u}{2} & 0\\
  -\frac{g \, v_d}{2} & \frac{g' \, v_d}{2} & 0 & -\frac{\lambda \, v_s}{\sqrt{2}} & - \frac{\lambda \, v_u}{\sqrt{2}} \\
    \frac{g \, v_u}{2} & -\frac{g' \, v_u}{2} & -\frac{\lambda \, v_s}{\sqrt{2}} & 0 & - \frac{\lambda \, v_d}{\sqrt{2}} \\
    0 & 0 & - \frac{\lambda \, v_u}{\sqrt{2}} &- \frac{\lambda \, v_d}{\sqrt{2}} & \frac{2 \, \kappa \, v_s}{\sqrt{2}}
  \end{array}\right).
  \label{neutralinomatrix}
\eea

To obtain the mass eigenstates, the neutralino mass matrix can be diagonalized
\bea
N^* \, M_{\chi^0} \, N^{-1} = diag(m_{\chi_1^0},m_{\chi_2^0},m_{\chi_3^0},m_{\chi_4^0},m_{\chi_5^0}),
\label{diagneutralino}
\eea 
where $m_{\chi_1^0} < m_{\chi_2^0} < m_{\chi_3^0} < m_{\chi_4^0} < m_{\chi_5^0}$. Then, the neutralino eigenstates can be written as $\chi_i^0=N_{i1}\tilde{B} + N_{i2}\tilde{W}_3 + N_{i3}\tilde{H}_u^0 + N_{i4}\tilde{H}_d^0 + N_{i5}\tilde{S}$, with the matrix $N$ defining the composition of the neutralinos.

In the chargino sector, the charged Higgsinos and the charged gaugino mix forming two couples of physical chargino $\chi_1^{\pm}$ and $\chi_2^{\pm}$. In the basis ($\tilde{W}^{\pm}$, $\tilde{H}_{d,u}^{\pm}$) the chargino mass matrix is given by
\bea
M_{\chi^{\pm}} = 
\left(\begin{array}{cc}
 M_2 & \sqrt{2} \, \cos \beta \, m_W \\
  \sqrt{2} \, \sin \beta \, m_W & \mu 
  \end{array}\right).
  \label{charginomatrix}
\eea

Using two unitary matrices the chargino mass matrix can be diagonalized to obtain the mass eigenstates
 \bea
U^* \, M_{\chi^{\pm}} \, V^{-1} = diag(m_{\chi_1^{\pm}},m_{\chi_2^{\pm}}),
\label{diagchargino}
\eea 
where $m_{\chi_1^{\pm}} < m_{\chi_2^{\pm}}$.

\section{Effective $\mu$-term, neutrino Yukawa values, and the Higgs sector}
\label{appendixB}

\begin{figure}[h!]
\begin{center}
 \begin{tabular}{cc}
 \hspace*{-10mm} \epsfig{file=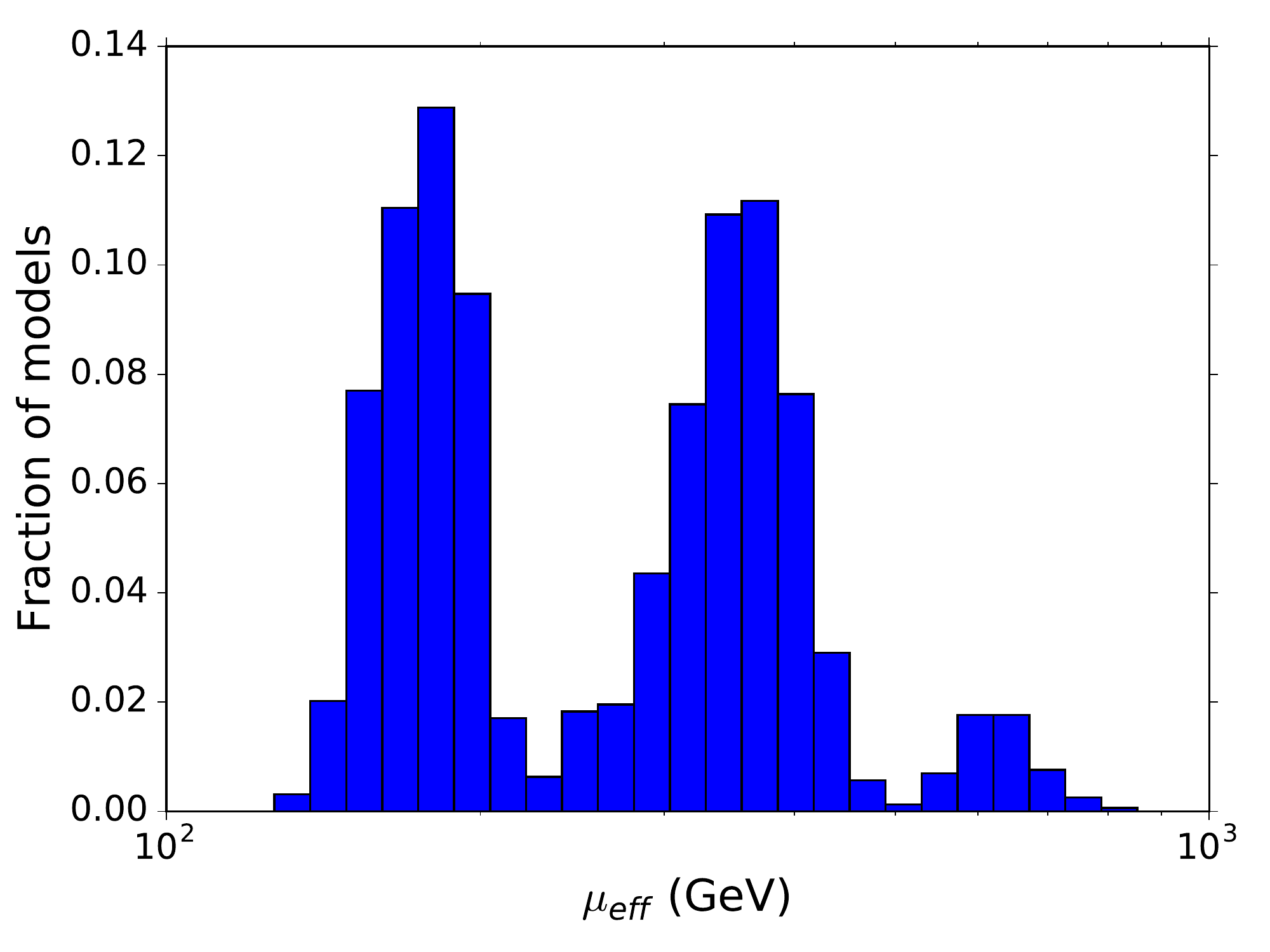,height=5.5cm} 
 \hspace*{-2mm}\epsfig{file=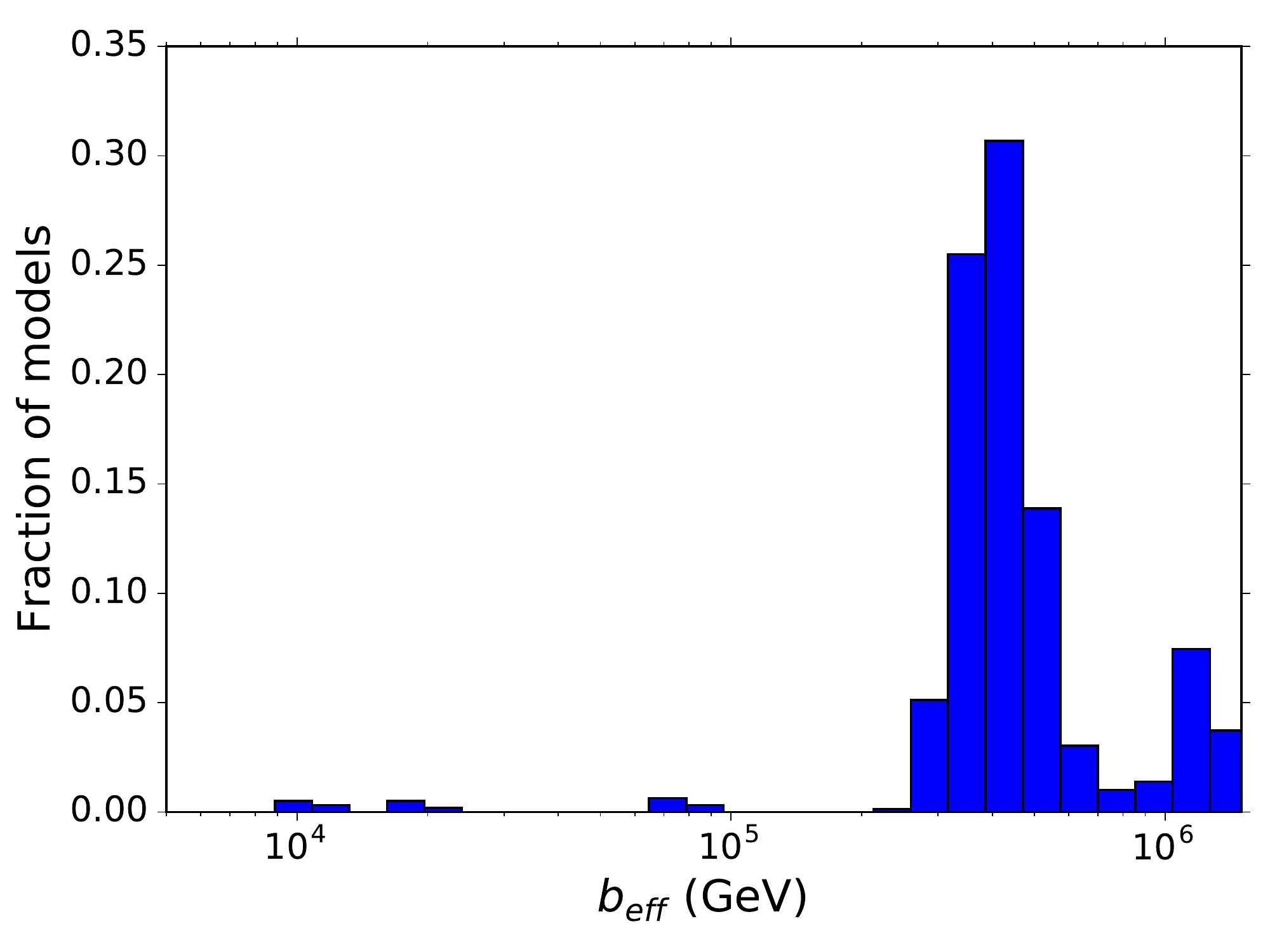,height=5.5cm} \\
        \hspace*{-10mm}\epsfig{file=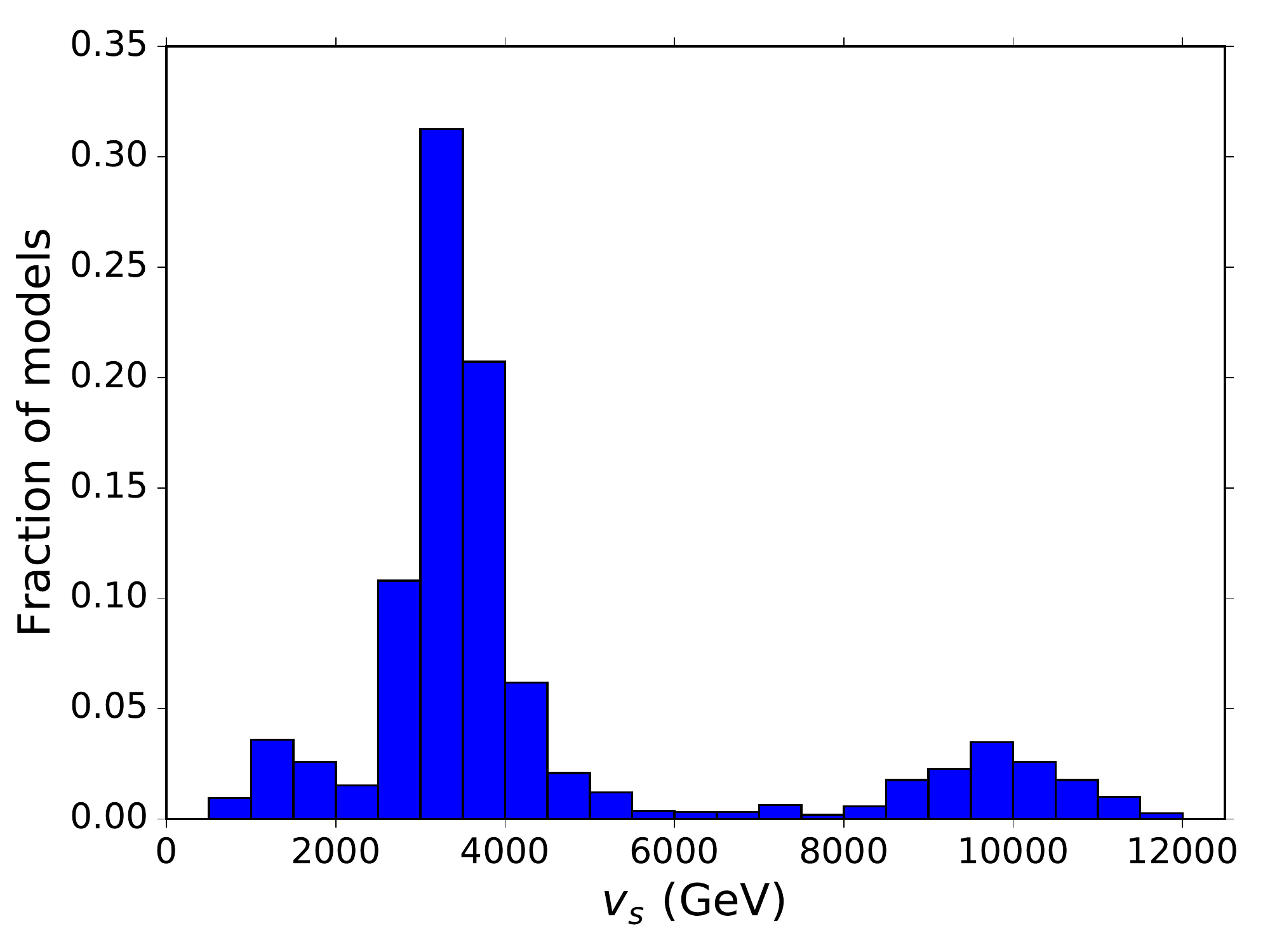,height=5.5cm}
        \hspace*{-2mm}\epsfig{file=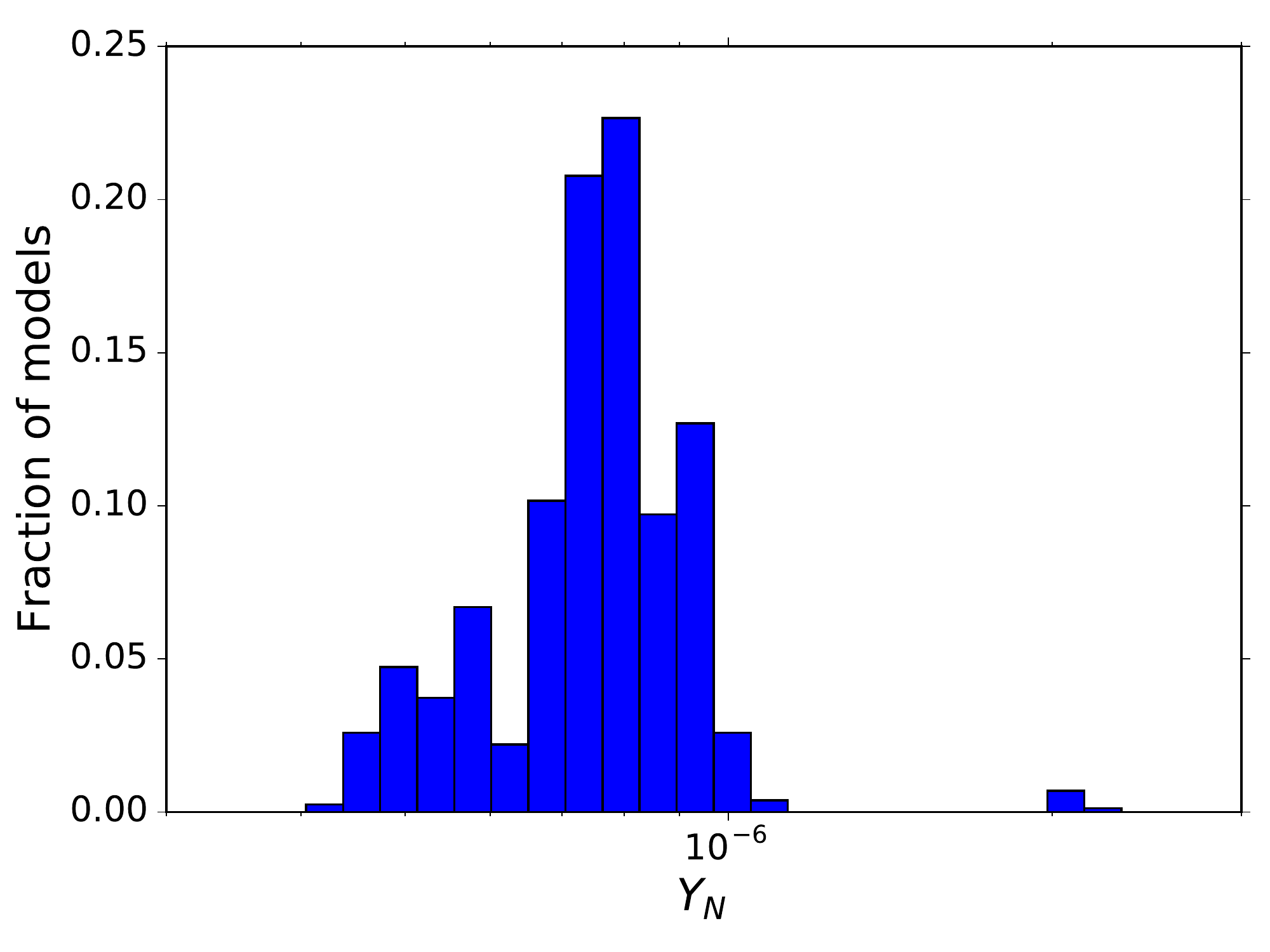,height=5.5cm} 
    \end{tabular}
    \captions{Values of $\mu_{eff}$, $b_{eff}$, $v_s$, and $Y_N$ ($m_{\nu}\sim O(0.1 \text{ eV})$) for the solutions that fulfill all the constraints considered in this work.}
    \label{histoAppendixB}
\end{center}
\end{figure}

In this appendix we show in Fig.~\ref{histoAppendixB} the values of $\mu_{eff}$, $b_{eff}$, $v_s$, and $Y_N$ for the solutions that fulfill all the constraints considered in this work. For the neutrino Yukawa couplings we obtain $m_{\nu}\sim O(0.1 \text{ eV})$.

\begin{figure}[h!]
\begin{center}
 \begin{tabular}{cc}\vspace*{-3mm}
 \hspace*{-12mm} \epsfig{file=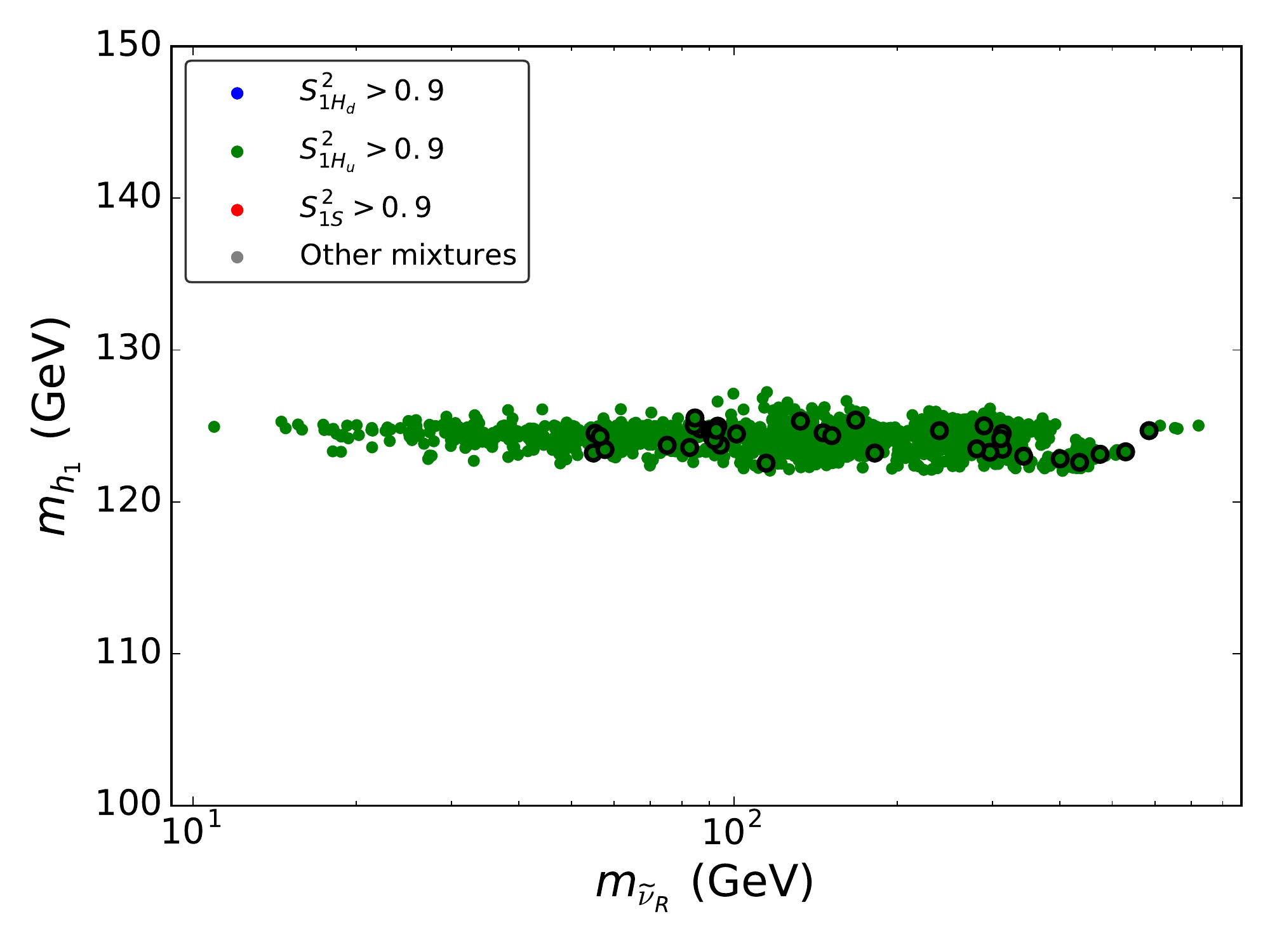,height=6.8cm} 
        \hspace*{-3mm}\epsfig{file=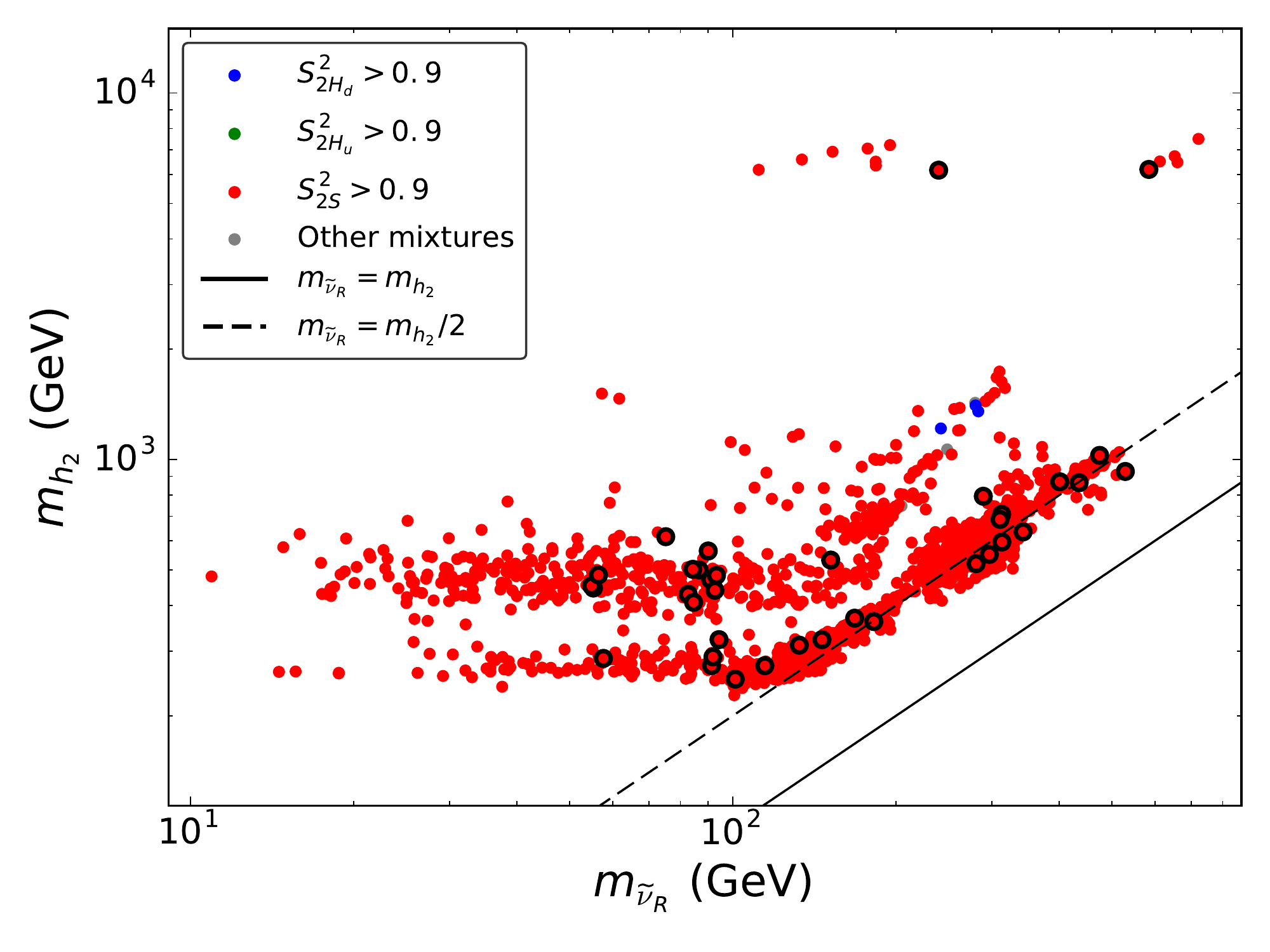,height=6.8cm}\\
        \hspace*{-12mm}\epsfig{file=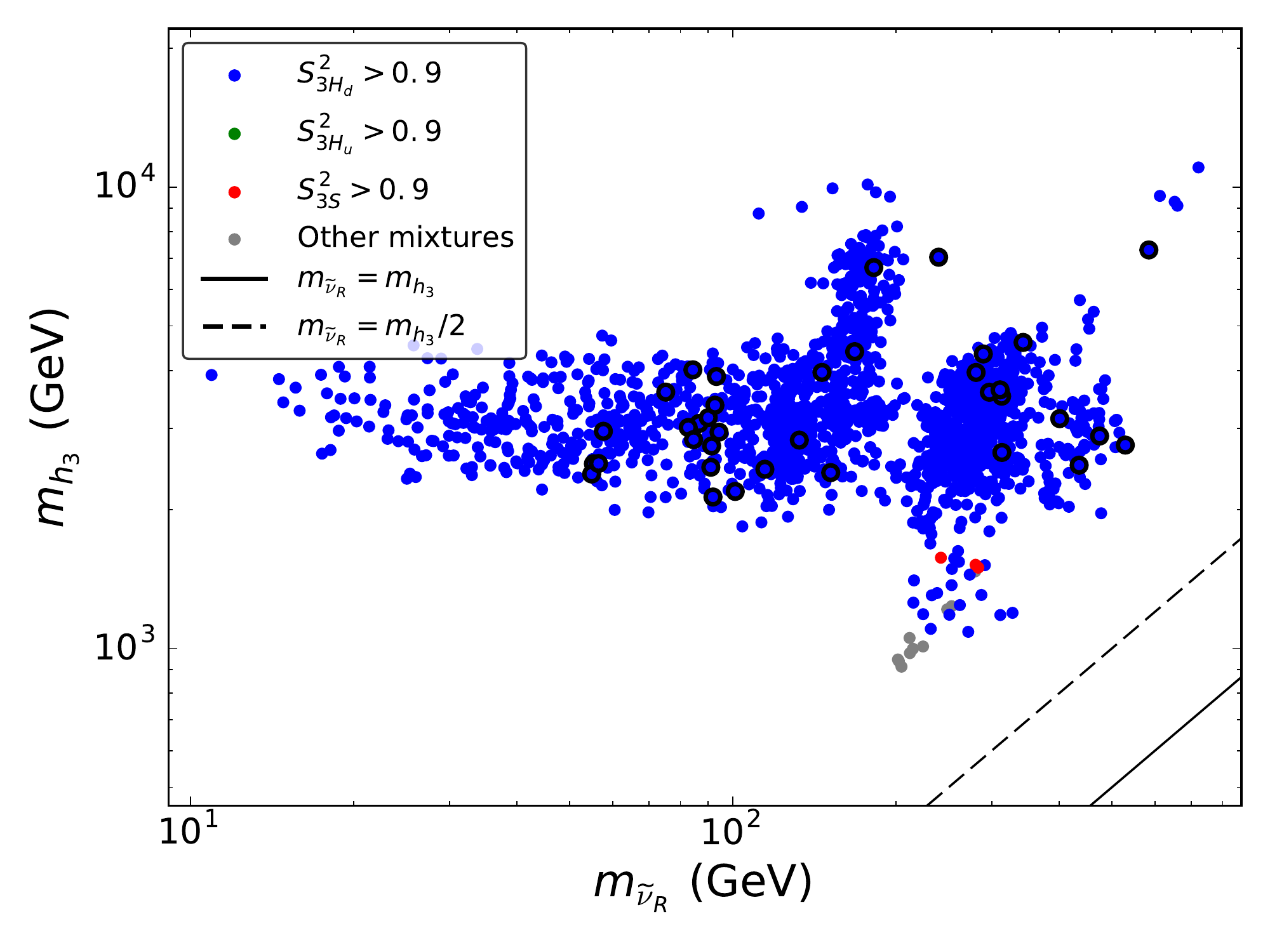,height=6.8cm} 
 \hspace*{-3mm} \epsfig{file=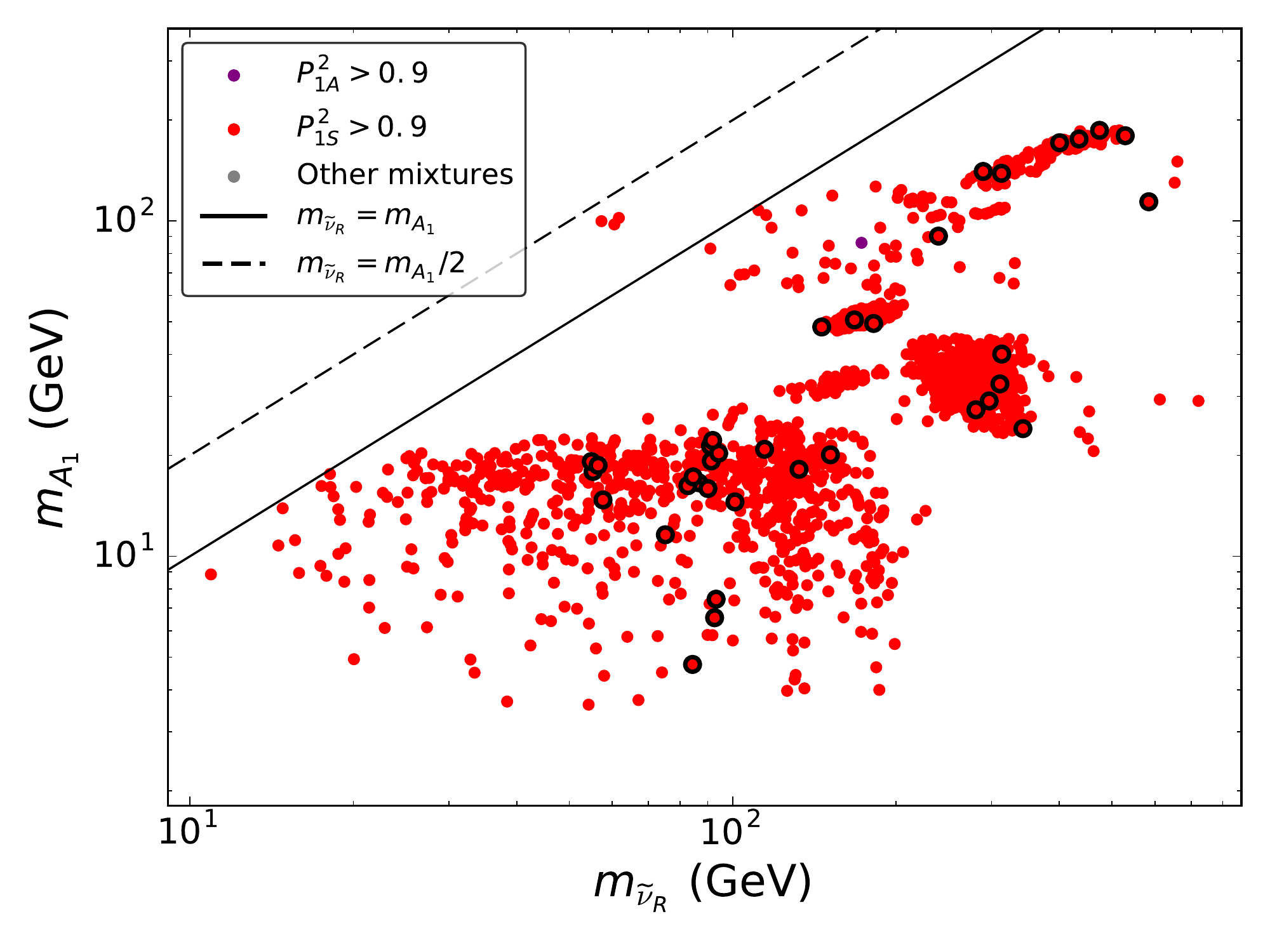,height=6.8cm} \\
     \end{tabular}
    \begin{tabular}{c}\vspace*{-3mm}
        \hspace*{-10mm}\vspace*{-3mm}\epsfig{file=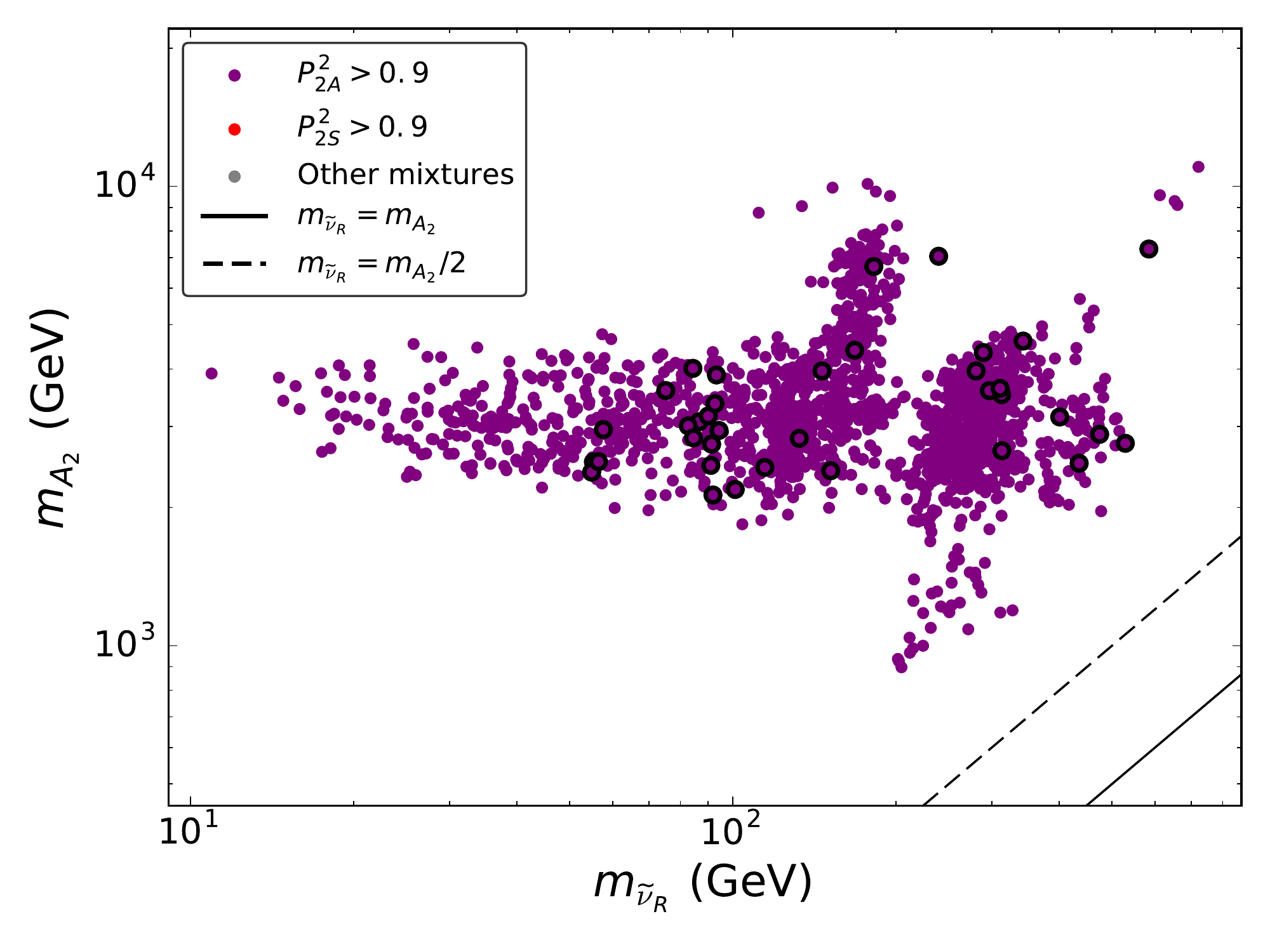,height=6.8cm} 
    \end{tabular}
    \captions{Masses of the CP-even and CP-odd scalars of the Higgs sector vs RH sneutrino mass for the solutions that fulfill all the constraints considered in this work. The color coding represents the dominant composition of each mass eigenstate (see Eq.~\ref{cphiggsmatrices}).}
    \label{higgssectorAppendixB}
\end{center}
\end{figure}

\begin{figure}[t!]
\begin{center}
 \begin{tabular}{c}
 \hspace*{-10mm}
 \epsfig{file=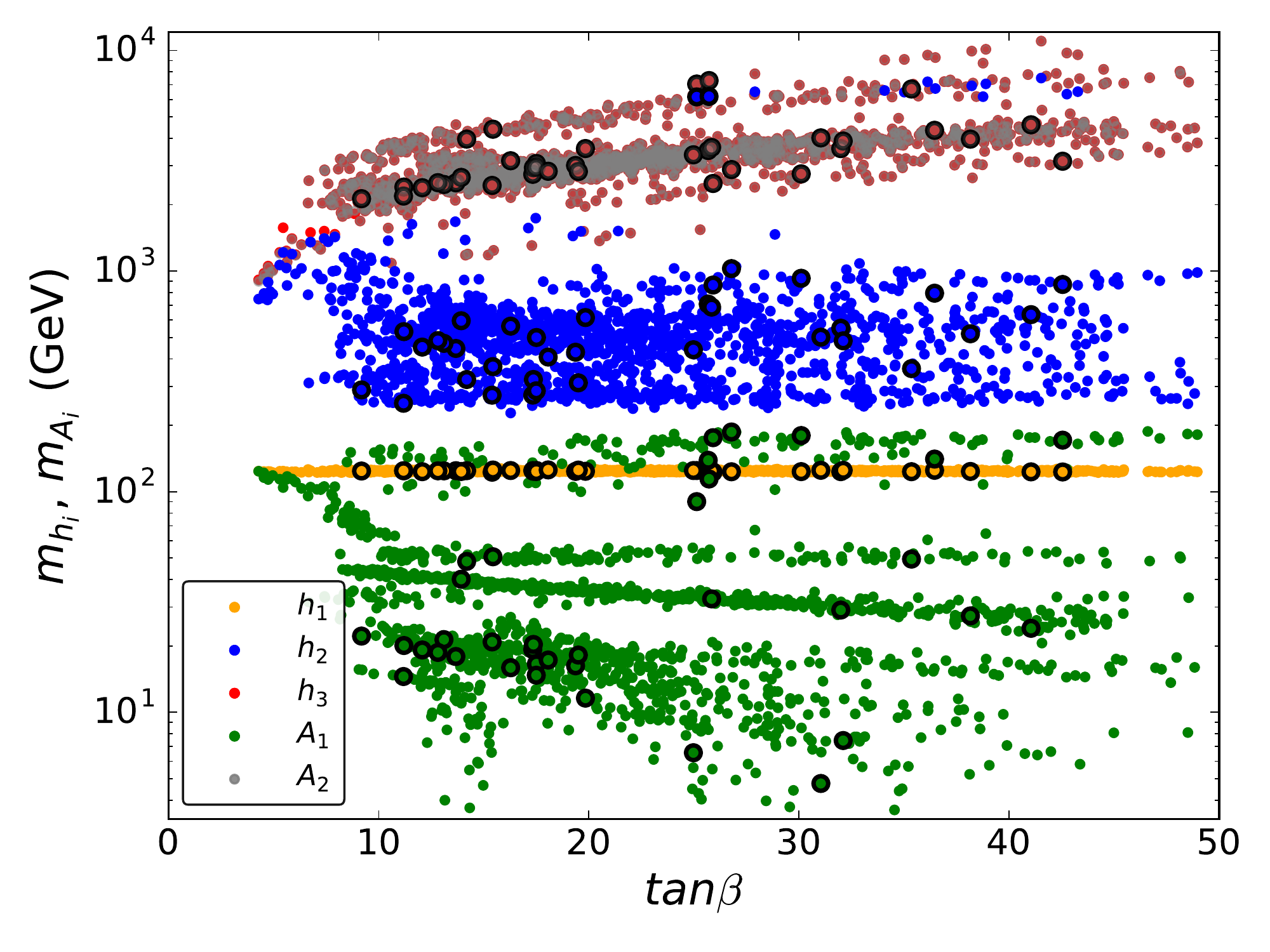,height=9cm} 
 
    \end{tabular}
    \captions{Masses of the Higgs sector vs $\tan \beta$. The color coding represents the mass eigenstates.}
    \label{higgssector2AppendixB}
\end{center}
\end{figure}

Regarding the Higgs sector, in Fig.~\ref{higgssectorAppendixB} we show the masses of the CP-even and CP-odd scalars and denote with color the dominant component of each mass eigenstate according to Eq.~\ref{cphiggsmatrices}. We can see that $h_1$ is always SM-like, $h_2$ and $A_1$ are singlet dominated for almost all solutions, and $h_3$ and $A_2$ are very heavy $\gtrsim 2$ TeV. Finally, to visualize the mass hierarchy in the Higgs sector, in Fig.~\ref{higgssector2AppendixB} we present the masses of the scalars as a function of $\tan \beta$. Notice that usually $m_{A_1} < m_{h_1} \sim 125 \text{ GeV } < m_{h_2} < m_{h_3} \simeq m_{A_2}$.

%%%%%%%%%%%%%%%%%%%%%%%%%%%%%%%%%%%%%%%%
%%%%%%%%%%%%%%%%%%%%%%%%%%%%%%%%%%%%%%%%

%\pagebreak

\bibliographystyle{utphys}
\bibliography{munussm}

\end{document}